\documentclass[aps,prb,reprint,nofootinbib]{revtex4-2}
\usepackage{amsmath,amssymb}
\usepackage{bm}

\usepackage{mathtools}

\usepackage[dvipdfmx]{graphics}
\usepackage{dcolumn}
\usepackage[dvipsnames]{xcolor}
\usepackage[caption=false,position=top,font=normalsize]{subfig}
\usepackage{here}

\usepackage{braket}
\usepackage[dvipdfmx, bookmarkstype=toc, colorlinks=true, linkcolor=cyan, pdfborder={0 0 0}, bookmarks=true, bookmarksnumbered=true]{hyperref}

\usepackage{hyperref}
\usepackage{tikz}
\usepackage{ulem}

\newcommand{\lr}[1]{\left({#1}\right)}

\newcommand{\lra}[1]{\left[{#1}\right]}

\newcommand{\D}[2]{\frac{d{#1}}{d{#2}}}

\newcommand{\pd}[2]{\frac{\partial#1}{\partial#2}}

\let\origlim\lim
\renewcommand{\lim}[2]{\origlim\limits_{{#1}\rightarrow{#2}}}

\newcommand{\Sum}[2]{\sum\limits_{#1}^{#2}}

\newcommand{\eps}{\epsilon}

\newcommand{\ua}{\uparrow}
\newcommand{\da}{\downarrow}

\begin{document}

\title{Nonlinear diode effect and Berezinskii-Kosterlitz-Thouless transition in purely two-dimensional noncentrosymmetric superconductors}

\author{Naratip Nunchot}
\author{Youichi Yanase}

\affiliation{Department of Physics, Graduate School of Science, Kyoto University, Kyoto 606-8502, Japan}

\date{\today}

\begin{abstract}

Phase diagrams and electronic transport properties of the helical states in purely two-dimensional (2D) Rashba superconductors coupled with in-plane Zeeman fields are studied. The continuum XY action is derived microscopically by integrating out the Gaussian amplitude fluctuation from the effective action. We show that the superfluid stiffness obtained from this procedure is exactly equivalent to the second-order derivative of the mean-field free energy density with respect to Cooper pair momentum, indicating an essential role of the amplitude fluctuation. The vortex core energy is also included in this work, and its effects on the Berezinskii-Kosterlitz-Thouless (BKT) transition line are discussed. The theory of nonlinear V-I characteristics in purely 2D superconductors is also revised to incorporate recent developments in the theory of the superconducting diode effect. The main results are as follows. We find that the nonlinear V-I characteristics of the system become nonreciprocal in finite in-plane Zeeman fields. This is reminiscent of the superconducting diode effect in 2D systems, although the critical current is zero in purely 2D superconductors. Furthermore, we find that the bare effective superfluid stiffness along the BKT transition line has a local minimum at a certain temperature, and the nonreciprocity of the V-I characteristics is strongly enhanced around this temperature.

\end{abstract}

\maketitle

\section{Introduction}

Nonreciprocal properties of superconductors have received much attention in recent years. In particular, the phenomenon that the critical currents of a superconductor become unequal in the positive and negative current directions, so-called \textit{the superconducting diode (SD) effect}, has been extensively investigated \cite{Ando, Daido1, Yuan, He-Nagaosa, Ilic, Daido2, Bauriedl, Narita, Nagaosa-Yanase}. This phenomenon is called \textit{intrinsic SD effect} when the nonreciprocity of critical currents arises from the properties of Cooper pairs in bulk \cite{Daido1}. In theoretical studies, the intrinsic SD effect has often been addressed by either the Bogoliubov-de Gennes (BdG) mean-field theory or the Ginzburg-Landau (GL) theory for mean-field two-dimensional (2D) superconductors. Since the nonreciprocity requires parity and time-reversal symmetry breaking, noncentrosymmetric superconductors under some time-reversal symmetry-breaking fields such as magnetic fields and ferromagnetic proximity effects are canonical platforms. Several mean-field theories considered the Fulde-Ferrell state (the helical state) of 2D Rashba superconductors in the in-plane Zeeman fields and showed the intrinsic SD effect \cite{Daido1, Yuan, Daido2, He-Nagaosa,Ilic}. A stripe phase of these models also shows the intrinsic SD effect \cite{Aoyama}, and the relationship between the Cooper pair momentum and supercurrent is similar to that in the helical phase when the supercurrent approaches the critical currents. In these models, the parity symmetry is broken by the Rashba spin-orbit coupling, and the supercurrent flowing perpendicular to the magnetic field shows the intrinsic SD effect. In another setup, where the Dzyaloshinsky-Moriya interaction breaks parity symmetry, the supercurrent parallel to the magnetic field causes the intrinsic SD effect, called the chiral SD effect \cite{Nunchot}.

In this paper, we address a natural question: How does the SD effect change in a purely 2D system? The long-range superconducting order is prohibited in 2D systems at finite temperatures, in contrast to three-dimensional and quasi-two-dimensional superconductors considered in the previous studies. In the classical XY model, the phase fluctuation destroys the off-diagonal long-range order, and the quasi-long-range order exhibits instead in the low-temperature region. The Berezinskii-Kosterlitz-Thouless (BKT) theory describes the transition that separates the quasi-long-range-ordered phase at low temperatures and the short-range-correlated phase at high temperatures~\cite{Kosterlitz1, Kosterlitz2}. The creation of bound vortex-antivortex pairs due to the quasi-long-range interaction is at the heart of this transition. In charge-neutral systems such as 2D superfluids, the phase fluctuation action coincides with the continuum XY model, and the BKT phase transition governs the critical phenomena~\cite{Nelson}. Thus, clarifying the effective action from fluctuation contributions is enormously important in studying the properties of the phase transition in 2D systems. In charged systems, the story is not simple. Although the vortex-antivortex pairs are also generated in charged systems, the interaction between them becomes short-range due to the electromagnetic (EM) shielding \cite{Pearl}, eliminating the phase transition. However, the transition can be treated as existing in 2D superconductors if the system size is smaller than or compatible with Pearl's length, which is usually longer than samples used in various experiments~\cite{Beasley}. 

In this paper, we study the BKT transition to the helical superconducting phase and the associated nonreciprocal transport in purely 2D superconductors. In previous studies, much progress has been made on the helical states in purely 2D fermionic superfluids. About one decade ago, the helical states of fermionic superfluids with the Rashba spin-orbit coupling or Zeeman coupling or both of them were studied with 2D ultracold atoms in mind~\cite{Gong, Yin, Xu2}. The continuum 2D-XY action for analyzing the BKT transition was microscopically derived by the Gaussian expansion with respect to the order parameter and the use of a unitary transformation \cite{Aitchison, Botelho}. On the Gaussian expansion of the action, the ultraviolet divergence inevitably appears after integrating out the spin wave part of phase fluctuations. This divergence is interpreted as a consequence of the zero-point energy of the Goldstone mode~\cite{Diener, Salasnich} and can be extinguished by the convergence factor regularization (CFR), which has been used in many works \cite{Gong, Yin, Xu2, Botelho, Diener, He, Bighin, Salasnich}, leaving only a finite Goldstone mode contribution to the action. Such a massless mode naturally appears in a neutral superfluid, but it should disappear in the Meissner phase of superconductors \cite{Anderson}. On the calculation of the BKT transition temperature, the Nelson-Kosterlitz (NK) relation \cite{Nelson} was extended to the case of anisotropic superfluid stiffness~\cite{Xu2}. In their work, the in-plane and out-plane magnetic field dependences of the BKT transition temperature were investigated, and the magnetic field-temperature ($h$-$t$) phase diagram, namely the BKT transition line, was obtained. In their calculations, the NK relation was approximated by neglecting the vortex core energy as in the traditional treatment.

Not only the properties of the superconducting phase transition but also the transport properties in purely 2D systems are significantly different from those in effectively three-dimensional systems. It is well known that supercurrent flow destroys the bound state of vortex-antivortex pairs in superfluid systems \cite{Ambegaokar}. This is also true for 2D superconductors, but there is a distinct feature not present in a neutral superfluid: nonlinear resistivity. When we neglect the finite size effect, the resistivity is strictly zero only in the zero current limit. On approaching the BKT transition temperature, a universal relation is satisfied such that the voltage across the sample becomes proportional to the cube of the supercurrent, $V\propto I^3$, in the low current limit~\cite{Halperin}. This expression has been utilized to identify the BKT transition in several experiments \cite{Hebard, Epstein, Fiory, Garland, Lin, Venditti, Saito, Hua, Liu, Weitzel}. Because the critical current disappears, the SD effect referring to the nonreciprocal critical current is absent in the purely 2D systems, as argued in Ref.~\cite{Nunchot}. However, the nonreciprocal response characteristic of the BKT transition is expected to occur. In fact, a phenomenological theory of purely 2D noncentrosymmetric superconductors, based on a simple free energy form, has studied the nonreciprocal resistivity at temperatures higher than the BKT transition temperature~\cite{Hoshino}. The treasure is then left open in the region below the BKT transition temperatures.

In this paper, we incorporate the recent developments in the intrinsic SD effect into the classical theory of the BKT transition for purely 2D superconductors and the theory of the associated V-I characteristics. In the study of the V-I characteristics, we focus on the region below the BKT transition temperature, where strong nonreciprocity is expected to appear. On obtaining the BKT transition line, the vortex core energy neglected in Ref.~\cite{Xu2} is included. In extending the theory of the V-I characteristics, Cooper pairs' momentum dependence of a set of parameters obtained in the mean-field theory is taken into account. We consider the helical states of a purely 2D Rashba superconductor coupled with in-plane Zeeman fields. The reason for considering the in-plane fields instead of the out-plane fields is that 2D superconductivity is fragile under the out-of-plane magnetic fields~\cite{Fiory, Garland, Hua, Qiu} due to the penetration of external magnetic flux \cite{DFisher}. There would be no superconducting transition in purely 2D superconductors under an out-plane field, especially in ideally clean samples \cite{DFisher}. The aim of the most is to show the nonreciprocal nonlinear V-I characteristics in the helical superconducting states. For this purpose, we simplify the analysis by considering the model in the clean limit and taking into account only the electrons near the Fermi surface as in the BCS theory.

As emphasized above, it is necessary to clarify the expression of the fluctuation action, at least at the Gaussian level. There are two issues involving this necessity. One is about the massless Goldstone mode contribution to the action appearing in superfluid systems as in Ref.~\cite{Xu2}. We will develop the procedure to extinguish such a contribution. The other is the formulation of the superfluid stiffness. Information on superfluid stiffness is needed to calculate the BKT transition line and nonlinear resistivity. It is argued that the superfluid stiffness should be obtained by differentiating $F(\bm{Q})$ with respect to $\bm{Q}$ to the second order, where $\bm{Q}$ is the Cooper pair center-of-mass momentum and $F(\bm{Q})$ is the free energy at the minimum for a given $\bm{Q}$ \cite{Liang}. This resolves the problem of the lattice geometry dependence of the superfluid stiffness in multi-sublattice systems \cite{Liang}. However, it is not obvious how the formula is derived in the quantum field theory framework used in Refs.~\cite{Gong, Yin, Xu2, Botelho, Diener}. In this paper, we use the functional integral method in a similar manner to these references and obtain an effective action. Then, we will show that after integrating out the Gaussian amplitude fluctuation, we reach the XY model where the superfluid stiffness is given by the second-order derivative of $F(\bm{Q})$ consistent with Ref.~\cite{Liang}.

The rest of this paper is organized as follows. After introducing the model in Sec.~\ref{II}, we show the analysis of the Gaussian fluctuations in Sec.~\ref{III}. In Sec.~\ref{IV}, we discuss the effective action derived by integrating out the phase and amplitude modes. We first show the renormalization of the superfluid stiffness from the amplitude fluctuation by integrating out the amplitude mode. Next, we relate the phase mode to the gauge field fluctuation. Then, we describe the prescription to manage the vortex fluctuation, that is, the vortex part of the continuum XY model, and review the BKT transition theory. In Sec.~\ref{5V}, the parameters to be numerically calculated are stated in the first place. After that, the numerical results on the properties of the zero-current states are reported and discussed. The behavior of the bare effective superfluid stiffness will be highlighted. In Sec.~\ref{VI}, the scheme to calculate the supercurrent is discussed. In Sec.~\ref{VII}, we review and improve the theory of V-I characteristics in purely 2D superconductors. In Sec.~\ref{VIII}, the numerical results on the V-I characteristics at various magnetic fields are shown and discussed in relation to the phase diagram. The nonlinear and nonreciprocal resistivity analogous to the SD effect is presented. At the end of this section, the finite-size effect is considered. We end this paper in Sec.~\ref{IX} with a summary and discussions. We adopt natural units $\hbar = k_{\mathrm{B}}=c=1$.

\section{Model\label{II}}

We start by defining a model and deriving its effective action. Let us consider a superconducting thin film lying in the $x_1x_2$-plane. The partition function $Z$ is given below:
\[      Z =  \int D[\Delta,\theta] \, e^{-S[\bar{\psi},\psi] },   \tag{1}   \]
\begin{align*}   
S[\bar{\psi},\psi] = \int & d^3x\,  \bigg[ \bar{\psi}_{a} \lr{ \partial_{\tau} + \frac{1}{2m} \hat{\bm{p}}^2 - \mu  } \psi_{a} \\
&+ \bar{\psi}_{a} \Big\{ h\,(\sigma_1)_{ab} + \alpha\bm{e}_3\cdot (\bm{\sigma}_{ab} \times \hat{\bm{p}})  \Big\} \, \psi_{b} \\
& - gA_{\mathrm{s}} \, \bar{\psi}_{\da}\bar{\psi}_{\ua}\psi_{\ua}\psi_{\da} \bigg],           \tag{2}    \label{eq2}     
\end{align*}
where $x=\{x_{\mu}\}_{\mu=0,1,2}=(x_0,\bm{x})$,~$\int d^3x = \int_0^{1/T}d\tau \int d^2\bm{x}$, $\tau=x_0$ is imaginary time, and $\hat{\bm{p}} = -i\nabla$. In Eq.~\eqref{eq2}, $A_{\mathrm{s}}$ is the area of the system, $\mu$ is the chemical potential which is equal to the Fermi energy $E_{\mathrm{F}}$, $g>0$ is a coupling constant of an attractive interaction, $T$ is a temperature, $h$ is the Zeeman field, $\alpha$ represents the Rashba spin-orbit coupling constant, and $\sigma_i$ is the $i$-th component of the Pauli matrices. Here, the Grassmann numbers $\bar{\psi},\psi$ denote fermion fields. By introducing auxiliary boson fields, namely $\Delta$ and its complex conjugate $\bar{\Delta}$, and performing the Hubbard-Stratonovich transformation, we reach the partition function $Z$ and the effective action $S_{\mathrm{eff}}$ as follows: 
\[      Z =  \int D[\Delta,\bar{\Delta}] \, e^{-S_{\textrm{eff}}[\Delta,\bar{\Delta}] },  \tag{3} \label{eq3}   \]
\[ S_{\textrm{eff}}[\Delta,\bar{\Delta}] = \int d^3x\, \frac{1}{gA_{\mathrm{s}}}| \Delta(x) |^2  - \frac{1}{2} \textrm{Tr} \ln G^{-1},    \tag{4}   \label{eq4}    \]
with
\[           G^{-1} =\begin{pmatrix}
\partial_{\tau} + H_{+} & -\hat{\Delta} \\
-\hat{\Delta}^{\dagger} & ~\partial_{\tau} + H_{-}
\end{pmatrix}.  \tag{5} \label{eq5} \]
Here, $\hat{\Delta} \equiv \Delta(x) \, (i\sigma_2)$, and $H_{\pm}$ are given as follows:
\[ H_{\pm} = \begin{bmatrix}
\pm\frac{1}{2m}\bm{\hat{p}}^2 \mp \mu & \alpha (\hat{p}_{2} \!\pm\! i\hat{p}_{1}) \! \pm \! h \\
\alpha (\hat{p}_{2} \! \mp \! i\hat{p}_{1}) \! \pm \! h & \pm \frac{1}{2m} \bm{\hat{p}}^2 \mp \mu 
\end{bmatrix}. \tag{6} \label{eq6} \]

Hereafter, a pair of a bosonic Matsubara frequency $\omega_n$ and momentum $\bm{p}$ is written as a generalized momentum $p=\{p_{\mu}\}_{\mu=0,1,2}=(p_0,\bm{p})$, where $p_0=-\omega_n$. Let us assume that the order parameter is in the form of the helical state, namely $\Delta(x)=\Delta(Q) \exp\,(iQ_{\mu}x_{\mu})$, where $Q_{\mu}$ is a generalized Cooper-pair momentum and $\Delta(Q)$ is a non-negative real number. In the expression of $\Delta(x)$, $Q_0$ is actually \textit{zero} due to the thermal equilibrium condition in the mean-field theory, but is introduced here for convenience in calculating some components of superfluid stiffness in Sec.~\ref{III}. Furthermore, we will see in Sec.~\ref{VI} that the generalized momentum $Q$, the order parameter of the helical state, cannot be classified from gauge fields. In that case, $Q_0\not=0$, but the summation between $Q_0$ and the 0-th component of the gauge fields becomes zero instead for the mean-field condition. For simplicity, we neglect the gauge field contribution except in Sec.~\ref{VI}. Introducing the $4\times4$ unitary matrix
\[ U = \begin{pmatrix}
e^{\frac{i}{2}Q_{\mu}x_{\mu} } & 0 \\
0 & e^{-\frac{i}{2}Q_{\mu}x_{\mu}}
\end{pmatrix},   \tag{7} \]
we can eliminate the factors $\exp\, (\pm iQ_{\mu}x_{\mu})$ in $G^{-1}$ making $\Delta$ independent of the coordinates. In this way,  $G^{-1}$ becomes $G^{-1}_0$ given by
\[           G_0^{-1} =\begin{pmatrix}
\partial_{\tau} + H_{\textrm{e0}} & -\hat{\Delta}_Q \\
-\hat{\Delta}^{\dagger}_Q & ~\partial_{\tau} + H_{\textrm{p0}}
\end{pmatrix},  \tag{8} \label{eq8} \]
at the mean-field level. Here,  $\hat{\Delta}_Q = \Delta(Q) \, (i\sigma_2)$, and $H_{\textrm{e0}}$ and $H_{\textrm{p0}}$ can be expressed as follows:
\begin{align*}
H_{\textrm{e0}} &= \begin{bmatrix}
\frac{1}{2m} \bm{\hat{p}}^2_+ -\mu +\! \frac{i}{2}Q_0 & \alpha (\hat{p}_{2+} \!+\! i\hat{p}_{1+}) \!+\! h \\
\alpha (\hat{p}_{2+} \!-\! i\hat{p}_{1+}) \!+\! h & \frac{1}{2m} \bm{\hat{p}}^2_+ -\mu +\! \frac{i}{2}Q_0 
\end{bmatrix}, \tag{9a} \label{eq9a} \\       
H_{\textrm{p0}} &= \begin{bmatrix}
-\frac{1}{2m} \bm{\hat{p}}^2_- + \mu - \! \frac{i}{2}Q_0 & \alpha (\hat{p}_{2-} \!-\! i\hat{p}_{1-}) \!-\! h \\
\alpha (\hat{p}_{2-}\!+\! i\hat{p}_{1-}) \!-\! h & -\frac{1}{2m} \bm{\hat{p}}^2_- + \mu - \! \frac{i}{2}Q_0
\end{bmatrix},      \tag{9b}  \label{eq9b}
 \end{align*}
where $\hat{p}_{i\pm} = \hat{p}_{i} \pm Q_i/2$. From Eqs.~(\ref{eq4}) and (\ref{eq8}), the mean-field free energy $F_{\textrm{mf}}$ is obtained as
\[  F_{\textrm{mf}}(Q) = -\frac{T}{2}\mathrm{Tr}\ln G^{-1}_0[\Delta(Q);Q] + \frac{1}{g}\lra{\Delta(Q)}^2. \tag{10} \label{eq10}            \]
For later convenience, here we introduce the functional
\[  F_{\textrm{mf}}[\Delta; Q] = -\frac{T}{2}\mathrm{Tr}\ln G^{-1}_0[\Delta;Q] + \frac{1}{g}|\Delta|^2. \tag{11} \label{eq11}            \]
 The above functional coincides with Eq.~(\ref{eq10}) when $\Delta = \Delta(Q)$. Note that the limit $Q_0\rightarrow0$ is taken after some successful calculations.
 
\section{Gaussian Fluctuation Action\label{III}}

In this section, we consider fluctuations of the order parameter. The theoretical formulation used here is similar to the previous work \cite{Yin}. There are two types of fluctuations: Phase fluctuation $\theta(x)$ and amplitude fluctuation $\varphi(x)$. By taking  into account these fluctuation fields, the order parameter is written as 
\[\Delta(x)=\lra{ \Delta(Q)+\varphi(x) } \exp \lra{ iQ_{\mu}x_{\mu} + i\theta(x) }.     \tag{12} \label{eq12} \] 
Substituting this into Eq.~(\ref{eq5}), $G^{-1}$ can be decomposed into the mean-field part $G_0^{-1}$ and the fluctuation part $\Sigma$, where $G^{-1}=G^{-1}_0 + \Sigma$. The mean-field part is given by Eq.~(\ref{eq8}), and the fluctuation part, derived by using Eq.~(\ref{eq8}), is obtained as follows:
\begin{align*}   
\Sigma =   i\varphi &\, (\sigma_2\otimes\sigma_1) \\
&+ \Big[ i\partial_{0}\theta + \frac{1}{4m}(\bm{Q}\cdot\nabla\theta)  + \frac{1}{8m}(\nabla\theta)^2     \Big] (\sigma_3 \otimes \sigma_0) \\
&-\frac{i}{4m} (\nabla^2\theta) + \frac{1}{2m}(\nabla\theta)\cdot\hat{\bm{p}} \\
&+ \frac{\alpha}{2} (\partial_2 \theta)\, (\sigma_3 \otimes \sigma_1) - \frac{\alpha}{2} (\partial_1 \theta)\, (\sigma_0 \otimes \sigma_2), \tag{13}
\end{align*}
where $\otimes$ denotes the tensor product.

To represent $\Sigma$ in momentum space, we insert the identity $\mathcal{I} \equiv V^{-1}\int d^3x\, \ket{x}\bra{x} =1$ to the left of $\Sigma$ obtaining the equation $\mathcal{I}\Sigma$, where $V=A/T$. Let $\braket{p|x} = \exp\,(ip_{\mu}x_{\mu})$, we have $\braket{p|p'} = \delta_{p,p'}$. Inserting $\bra{p}$ and $\ket{p'}$ into the left and right of $\Sigma$, respectively, we obtain $\Sigma\,(p,p') = \bra{p}\mathcal{I}\Sigma \ket{p'}$ as the $(p,p')$ element of $\Sigma$. The integration over real space in the equation can be removed by performing the Fourier transformation of each fluctuation field. There, it must be noted that the results depend on the boundary conditions. For the amplitude field $\varphi$, the periodic boundary condition is imposed. For the phase field $\theta$, it is, however, appropriate to impose only the periodic boundary condition on imaginary time because, in the below, we have to work on a situation in which the phase difference between two ends of the system is imposed to be linear in the distance of those two ends. In this situation, the summation of the spatial-momentum Fourier components of $\theta$ cannot converge to $\theta$ at the edges. To overcome this problem, we assume that $\theta\sim \eps_{\mu} x_{\mu}$ at the edges. The $\theta$ terms in $\Sigma$ are differentiated not less than one time, and the periodic boundary conditions in space are recovered for $\partial_i\theta$. Therefore, instead of $\theta$ we perform the Fourier transformation on its gradient, $a_{\mu}\equiv \partial_{\mu}\theta$. In this way, $\Sigma\,(p,p')$ is calculated as follows:

\[ \Sigma\, (p,p') =  \Sigma_1\,(p,p') + \Sigma_2\,(p,p'), \tag{14} \]
\begin{align*}   
\Sigma_1(p,p') &= \Sum{k}{} \delta_{p-p',k}\,  \bigg\{ i\varphi_{k} \, (\sigma_2\otimes\sigma_1) \\
&\quad + \Big[ ia_{0k} + \frac{1}{4m}(\bm{Q}\cdot\bm{a}_k)   \Big] (\sigma_3 \otimes \sigma_0) \\
&\quad + \frac{1}{4m} (\bm{k}\cdot\bm{a}_k) + \frac{1}{2m}(\bm{a}_k\cdot\bm{p}')  \\
&\quad + \frac{\alpha}{2} a_{2k}\, (\sigma_3 \otimes \sigma_1) - \frac{\alpha}{2} a_{1k}\, (\sigma_0 \otimes \sigma_2) \bigg\}, \tag{15a}
\end{align*}
\[\Sigma_2(p,p') = \frac{1}{8m}\Sum{k,k'}{} \delta_{p-p',k'-k} \, (\bm{a}_{k}\cdot\bm{a}_{-k'})\, (\sigma_3 \otimes \sigma_0).    \tag{15b} \]

Next, we expand $\mathrm{Tr}\ln\, G^{-1}$ to the second order of fluctuation fields. That is
\[ \mathrm{Tr} \ln G^{-1} = \mathrm{Tr} \ln G^{-1}_0 + \mathrm{Tr} \lra{ G_0\Sigma } - \frac{1}{2}\mathrm{Tr} \lra{G_0\Sigma_1 G_0\Sigma_1}. \tag{16} \label{eq16} \]
The second term on the r.h.s. of the above equation can be expressed as below:
\begin{align*}
\mathrm{Tr}\, \lra{G_0\Sigma} &= \Sum{p}{}\mathrm{tr}\lra{ G_0(p)\Sigma(p,p)} \\ 
&= \mathcal{R}_1 + \Big( \Sum{p}{}\mathrm{tr} \, G(p)\,(\sigma_3 \otimes \sigma_0)\Big) \Sum{k}{}\, \bm{a}_k\cdot\bm{a}_{-k}\\
&= \mathcal{R}_1 + R_0V^{-1} \int d^3x \, (\nabla\theta)^2, \tag{17} \label{eq17}
\end{align*}
where $\mathcal{R}_1$ is linear in $a_{\mu k}$ and $\varphi_k$, $R_0 = \sum_{p} \mathrm{tr} \, G(p)\,(\sigma_3 \otimes \sigma_0)$, and $\mathrm{tr}$ denotes the trace of a matrix for a given $p$. The first-order terms of $a_{\mu k}$ are nonzero if a constrained condition such that a constant supercurrent flows through the system is imposed [see Sec.~\ref{VI}]. This $\mathcal{R}_1$ term does not appear in Refs.~\cite{Gong, Yin, Xu2}, and indeed it determines a helical state under a given supercurrent. By a method mentioned in Sec.~\ref{VI}, the relation between the Cooper pair momentum $\bm{Q}$ and supercurrent can be found without knowing the explicit expression of $\mathcal{R}_1$ in terms of $G(p)$. Moreover, as also explained there, the action due to the $a_{\mu k}$ terms in the $\mathcal{R}_1$ term is offset by the supercurrent source action term, resulting in the mean-field condition of a superconductor under a supercurrent. The source term will be introduced in Sec.~\ref{VI}. The $\varphi_k$ term in the $\mathcal{R}_1$ term is related to the gap equation and thus vanishes. The third term on the r.h.s. of Eq.~(\ref{eq16}) can be expressed as below:

\begin{widetext}
\begin{align*}
& \mathrm{Tr} \lra{G_0\Sigma_1G_0\Sigma_1} = \Sum{p,p'}{}\mathrm{tr} \lra{G_0(p) \Sigma_1(p,p')G_0(p') \Sigma_1(p',p)}   \\
&\simeq 2 \int d^3x \Big[ (D - 2g^{-1})\, \varphi^2 + D_{\mu\nu}  (\partial_{\mu} \varphi)(\partial_{\nu} \varphi) +E_{\mu\nu} (\partial_{\mu} \varphi)(\partial_{\nu} \theta) + 2\, C_{\mu} \varphi \, (\partial_{\mu}\theta) + \lr{J_{\mu\nu} +  \delta_{\mu\nu} R_0V^{-1}} (\partial_{\mu} \theta) (\partial_{\nu} \theta) \Big], \tag{18} \label{eq18}
\end{align*}
\end{widetext}
where higher-order derivative terms are neglected, and the coefficients of each fluctuation term are independent of the fluctuation fields and coordinates. Note that there is also the term $g^{-1}(2\varphi\,\Delta(Q) + \varphi^2)$ in Eq.~(\ref{eq4}), where the $\varphi$-linear term is responsible for the gap equation. From this and  Eqs.~(\ref{eq17}) and (\ref{eq18}), the Gaussian fluctuation action $S_{\textrm{fluc}}$ is obtained as

\begin{align*}    
S_{\textrm{fluc}} =  \frac{1}{2} \int d^3x & \Big[ D \varphi^2 + D_{\mu\nu}  (\partial_{\mu} \varphi)(\partial_{\nu} \varphi) \\
&+E_{\mu\nu} (\partial_{\mu} \varphi)(\partial_{\nu} \theta) + 2\, C_{\mu} \varphi \, (\partial_{\mu}\theta) \\
&+ J_{\mu\nu} (\partial_{\mu} \theta) (\partial_{\nu} \theta) \Big]. \tag{19} \label{eq19}
\end{align*}

Now, let us consider expressions of each coefficient in the above equation. It will become clear later that the terms involving a derivative of $\varphi$ are not important to the following discussion. Thus, we only consider the expressions of coefficients $D,~C_{\mu}$, and $J_{\mu\nu}$. Although they can be obtained by carefully expanding Eq.~(\ref{eq18}), one can get an expression straightforwardly by relating each term to the mean-field free energy $F_{\textrm{mf}}$, which is more convenient for the numerical calculation. To do this, note that $TS_{\mathrm{fluc}}$ can be regarded as a change in the mean-field free energy, Eq.~(\ref{eq10}), to the second order of increments for given values of $\varphi$ and $\theta$. Let $\varphi = \eps$ and $\theta = u_{\mu}x_{\mu}$ where $\eps$ and $u_{\mu}$ are positive infinitesimal values. By using them to compute the partial derivatives of Eq.~(\ref{eq11}) at $\Delta = \Delta(Q)$, it is apparent that the second-order partial derivatives of Eq.~(\ref{eq11}) at $\Delta = \Delta(Q)$ contain only the contributions of Eq.~(\ref{eq18}). Therefore, $D,~C_{\mu}$, and $J_{\mu\nu}$ are given as follows:

\begin{align*}
D& = \lr{\frac{\partial^2 }{\partial \Delta^2}  f_{\textrm{mf}}[\Delta;Q]     }_{\Delta=\Delta(Q)}, \tag{20a} \label{eq20a} \\
C_{\mu} &=  \lr{\frac{\partial^2 }{\partial \Delta \, \partial Q_{\mu} }  f_{\textrm{mf}}[\Delta;Q]        }_{\Delta=\Delta(Q)}, \tag{20b} \label{eq20b} \\
J_{\mu\nu} &=  \lr{ \frac{\partial^2 }{\partial Q_{\mu}  \partial Q_{\nu} }   f_{\textrm{mf}}[\Delta;Q] }_{\Delta=\Delta(Q)},     \tag{20c} \label{eq20c} 
\end{align*}
where $f_{\textrm{mf}}[\Delta;Q]=F_{\textrm{mf}}[\Delta;Q]/A_{\mathrm{s}}$. By invoking the $Q_0$ term in Eqs.~(\ref{eq9a}) and (\ref{eq9b}), we can immediately know that $C_0$ is a purely imaginary number. This can be shown by replacing $iQ_0$ with a real number $\omega$ in Eq.~(\ref{eq10}) or Eq.~(\ref{eq11}). This replacement confirms that the mean-field free energy $F_{\mathrm{mf}}[\Delta;Q]$ is a real number. Thus, its first-order derivative with respect to $-i\omega$ is a purely imaginary number. Moreover, by looking at the form of $G^{-1}_0[\Delta;Q]$ in Eq.~(\ref{eq11}), it is shown that $C_1=0$ in our model if $Q_1=0$.

\section{Total Superfluid Stiffness\label{IV}}

Here, we integrate out the amplitude field $\varphi$ in Eq.~(\ref{eq19}). To do this, the measure of the path integral in Eq.~(\ref{eq3}) has to be clarified. Substituting Eq.~(\ref{eq12}) into the measure in Eq.~(\ref{eq3}) and calculating its Jacobian, the measure changes from $D[\Delta,\bar{\Delta}] = \prod d\Delta \, d\bar{\Delta}$ to $D[\varphi,\theta] = \prod \,2 (\Delta(Q)+\varphi)\, d\varphi\, d\theta$. The extra $\varphi$ in $D[\varphi,\theta]$ makes the analysis difficult. From now on, we will neglect such a term, and $D[\varphi,\theta]$ is approximated as $\prod 2 \Delta(Q)\,d\varphi\, d\theta$. Indeed, to set the Jacobian equal to 1, we can use
\[\Delta(x) = \sqrt{\Delta^2(Q)+\eta(x)} \exp \lra{ iQ_{\mu}x_{\mu}+ i\theta(x) },       \tag{21}     \label{eq21}\] 
together with the approximation $\varphi(x) = \sqrt{\Delta^2(Q)+\eta(x)}\!-\!\Delta(Q)\! \simeq \! \eta(x)\,[2\Delta(Q)]^{-1}$. Using them, one can reach an expression equivalent to Eq.~(\ref{eq19}).

From the above consideration, the partition function of the system becomes
\[    Z =   e^{ -F_{\mathrm{mf}}(Q)/T }\int D[\varphi,\theta] \, e^{ -S_{\mathrm{I}}[\partial\theta] - S_{\mathrm{fluc}}[\varphi,\partial\theta]  }, \tag{22} \label{eq22}  \]
where $\partial=\{\partial_{\mu}\}_{\mu=0,1,2}$ and $S_{\mathrm{I}}[\partial\theta]$ denotes the contribution from the linear term in $\partial_{\mu}\theta$ stemmed from the $\mathcal{R}_1$ term in Eq.~(\ref{eq17}). To perform the integration with respect to $\varphi$, it is better to work in momentum space. Extracting the terms coupled with $\varphi$ from Eq.~(\ref{eq19}) as the action $S_{\varphi\theta}$, we obtain:
\begin{widetext}
\begin{align*}
S_{\varphi\theta} &=  \frac{1}{2}\Sum{p}{} \Big[ \mathcal{K}_{0p} \varphi_p \varphi_{-p} - iE_{\mu\nu} p_{\mu} a_{\nu p} \varphi_{-p} + 2\, C_{\mu} a_{\mu p}  \varphi_{-p} \Big] \\
&=\frac{1}{2}\Sum{p}{} \Big[ \mathcal{K}_{0p} \tilde{\varphi}_p \tilde{\varphi}_{-p}  +  \frac{1}{4} \mathcal{K}^{-1}_{0p}\big( iE_{\mu\nu}p_{\mu}a_{\nu p}  -  2\, C_{\mu}a_{\mu p}    \big) \big( iE_{\alpha\beta}p_{\alpha}a_{\beta, -p}  +  2\, C_{\alpha}a_{\alpha, -p}    \big)\Big], \tag{23} \label{eq23}
\end{align*}
where $\mathcal{K}_{0p}=D  + D_{\mu\nu}  p_{\mu}p_{\nu}$ and $\tilde{\varphi}_{p} = \varphi_{p} - \frac{1}{2}\mathcal{K}^{-1}_{0p}\big( iE_{\mu\nu}p_{\mu}a_{\nu p}  -  C_{\mu}a_{\mu p}    \big) $. Now, the action has become a perfect square. Inserting this into Eq.~(\ref{eq22}), we perform the Gaussian integral. Then, we obtain the new effective action of the fluctuation part as follows:
\begin{align*}        
S_{\mathrm{fluc,eff}} &=  \frac{1}{2} \int d^3x \, J_{\mu\nu} (\partial_{\mu} \theta) (\partial_{\nu} \theta) + \frac{1}{2}\Sum{p}{} \ln \, \lr{1-e^{-E_{\textrm{amp}}(p)/T }} \\
&\quad -\frac{1}{8}\Sum{p}{} \mathcal{K}^{-1}_{0p} \big[ E_{\nu\mu}E_{\alpha\beta} \, p_{\nu} p_{\alpha}  + 2i E_{\alpha\beta} C_{\mu}p_{\alpha}  - 2i E_{\alpha\mu} C_{\beta}p_{\alpha}   + 4\, C_{\mu} C_{\beta} \big] a_{\mu p} a_{\beta, -p}.     \tag{24}  \label{eq24}             
\end{align*}
\end{widetext}
Here, $E_{\textrm{amp}}(p)$ is the excitation energy of amplitude modes, and we used the CFR in the second term of the r.h.s. of the above equation. A comprehensive explanation of this regularization scheme is given in Refs.~\cite{Diener} and \cite{Salasnich}. Hereafter, this term is assumed to be smaller than the others in the range of the parameters to be examined and is neglected. The assumption should be valid as long as the supercurrent in the system does not reach its critical value. Moreover, since we are interested only in the terms with differential order less than 2, we will keep only $O(1)$ terms in the second row of the above equation. Therefore, we obtain the effective fluctuation action $S_{\mathrm{fluc,eff}}$ as a functional of $\partial\theta$ as follows:

\[  S_{\mathrm{fluc,eff}} \simeq  \frac{1}{2} \int d^3x \,  \tilde{J}_{\mu\nu} (\partial_{\mu} \theta) (\partial_{\nu} \theta),  \tag{25} \label{eq25}    \]
where $\tilde{J}_{\mu\nu}$ is the $\mu\nu$ component of the total superfluid stiffness and is given as below: 
\[     \tilde{J}_{\mu\nu} =      J_{\mu\nu} - \frac{C_{\mu}C_{\nu}}{D}.     \tag{26} \label{eq26}          \]

We now show that the above equation can be deformed to an equation in which $F_{\textrm{mf}}[\Delta; Q]$ in Eq.~(\ref{eq20c}) is replaced by $F_{\textrm{mf}}(Q)$. The gap equation tells us that $\partial_{\Delta} F_{\textrm{mf}}[\Delta; Q]=0$. Let us regard it as an equation of state $U(\Delta;Q)=\partial_{\Delta} F_{\textrm{mf}}[\Delta; Q]$. Then, one reaches $dU = 0 = U_{\Delta}d\Delta + U_{Q_{\mu}}dQ_{\mu}$. Now, regarding $\Delta$ as a function of $Q$, we obtain the following:
\[   \pd{\Delta(Q)}{Q_{\mu}} = -\lra{ \frac{ \frac{\partial^2 }{\partial \Delta \, \partial Q_{\mu} }  f_{\textrm{mf}}[\Delta;Q]         }{   \frac{\partial^2 }{\partial \Delta^2}  f_{\textrm{mf}}[\Delta;Q]     } }_{\Delta=\Delta(Q)}   = -\frac{C_{\mu}}{D}. \tag{27} \label{eq27}    \]
Using the above equation in Eq.~(\ref{eq26}), $\tilde{J}_{\mu\nu}$ can be rewritten as below:
\[           \tilde{J}_{\mu\nu} =  \frac{\partial^2 }{\partial Q_{\mu}  \partial Q_{\nu} }   f_{\textrm{mf}}(Q), \tag{28} \label{eq28}       \]
where  $f_{\textrm{mf}}(Q)= F_{\textrm{mf}}(Q)/A_{\mathrm{s}}$. Thus, the statement at the beginning of this paragraph is true. Note that $Q_0$ is brought to be zero after performing the derivative in the above equation. Moreover, as seen from Eqs.~(\ref{eq9a}) and (\ref{eq9b}), the derivative with respect to $Q_0$ at $Q_0=0$ can be thought of the derivative with respect to $2i\mu$ at $\mu=E_{\mathrm{F}}$ with $Q_0=0$.

\captionsetup{font=normal,justification=raggedright,singlelinecheck=false}

\begin{figure}[b]
\centering

\includegraphics[width=5cm]{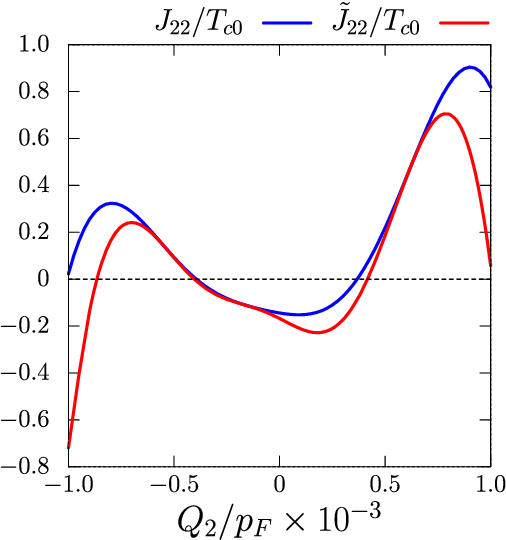}
\caption{The momentum $Q_2$ dependence of the $x_2$-component of the superfluid stiffness. The blue and red lines show $J_{22}/T_{\mathrm{c0}}$ and the total superfluid stiffness $\tilde{J}_{22}/T_{\mathrm{c0}}$, respectively. The reduced temperature $t \equiv T/T_{\mathrm{c0}}=0.6$ and the magnetic field $h/T_{\mathrm{c0}}=1.57$ lie on the BKT transition line in Fig.~\ref{fig3a}. $T_{\mathrm{c0}}$ is the mean-field transition temperature at zero magnetic field.}

\label{fig1}
\end{figure}

Here we mention two related works~\cite{Boyack, Samokhin} for the above expression. In Ref.~\cite{Boyack}, the amplitude fluctuation is integrated out from the action and renormalized to the superfluid stiffness. However, the simple form like Eq.~(\ref{eq28}) was not derived, and the helical state was not investigated. In Ref.~\cite{Samokhin}, the stability of the GL free energy for a 1D helical superconductor was examined, and expressions similar to Eqs.~(\ref{eq27}) and (\ref{eq28}) were derived. However, the connection of those quantities to the superfluid stiffness and the effective action for the Gaussian fluctuation were not studied there. Moreover, the formalism used there is different from ours, and the questions remained why the superfluid stiffness should be given by Eq.~(\ref{eq28}) and how to obtain the expression based on the functional integral approach in quantum field theories. Our approach succeeds in answering these questions in the following way. Since the coefficient given by Eq.~(\ref{eq28}) is coupled to the Gaussian phase fluctuation as in Eq.~(\ref{eq25}), Eq.~(\ref{eq28}) has thus to be interpreted as a superfluid stiffness. Importantly, as seen in the procedures to obtain Eqs.~(\ref{eq24}) and (\ref{eq26}), Eq.~(\ref{eq28}) results from integrating out the uniform amplitude fluctuations from the fluctuation action.

An example of $J_{22}$ and $\tilde{J}_{22}$ curves as functions of a Cooper pair momentum $Q_2$ is shown in Fig.~\ref{fig1}. The mean-field free energy [or the condensation energy defined in Sec.~\ref{5V A}] used in obtaining these curves are obtained by the formulation and parameter values stated in Sec.~\ref{5V A}. The total superfluid stiffness $\tilde{J}_{22}$ renormalized by amplitude fluctuations exactly coincides with the second-order derivative of the mean-field free energy density in Fig.~\ref{fig6c}. The $J_{22}$ without amplitude fluctuations is almost equivalent to $\tilde{J}_{22}$ for the Cooper pair momentum which minimizes the mean-field free energy, $Q_2/p_{\rm F} \sim 0.7 \times 10^{-3}$, where $p_{\rm F}$ is a Fermi momentum satisfying $p^2_{\mathrm{F}}=2mE_{\mathrm{F}}$. However, we see significant deviation indicating the essential role of amplitude fluctuations in the supercurrent-carrying helical state. Moreover, as promptly observed from the figure, the above $Q_2$ minimizing the free energy does not coincide with the Cooper pair momentum that maximizes $J_{22}$ and $\tilde{J}_{22}$. This issue will be addressed in Sec.~\ref{5V C}. Although we do not show the results of $J_{11}$ and $\tilde{J}_{11}$ here, we checked that their values differ by less than $0.1\%$ when $J_{22}$ is non-negative and $Q_1$ is zero. This is expected because from the discussion below Eq.~(\ref{eq20c}), $C_1=0$ at $Q_1=0$, and thus, $\tilde{J}_{11}=J_{11}$ when $Q_1=0$. A small deviation in the numerical results is expected to be due to rounding errors.

It should be noted that the superfluid stiffness given by Eq.~(\ref{eq28}) is not gauge-invariant. To obtain a result retaining the gauge invariance, longitudinal gauge fluctuation has to be further integrated. Although we plan to discuss it in Sec.~\ref{IV A}, the correction from this procedure is ultimately neglected in the numerical calculations in Secs.~\ref{5V} and \ref{VIII} for the reasons described in the next section.

\subsection{Gauge Field Fluctuation\label{IV A}}

The form of the effective fluctuation action Eq.~(\ref{eq25}) implies a Goldstone mode. In this section, we show that the Goldstone mode associated with the spin wave part disappears when the gauge field fluctuation is included in the theory. In the presence of the gauge field fluctuation, the phase fluctuation field $\theta$ in Eqs.~(\ref{eq22}) and (\ref{eq25}) is now written as
$\theta=\theta_r+\theta_a$, where $\theta_r$ and $\theta_a$ denote the phase part and the gauge field part, respectively. The phase part can be furthermore classified into the spin wave part $\theta_s$ and the vortex part $\theta_v$. The gauge field part is represented as $\theta_a = 2e\int^x\, dx'_{\mu}A_{\mu}(x')$, where $A(x')=\{A_{\mu}(x')\}_{\mu=0,1,2}$ denotes a fluctuating gauge field. Following this, the partition function $Z$ in Eq.~(\ref{eq22}) becomes a functional of the gauge field $A$, namely $Z=Z[A]$. To obtain the total partition function, we have to include the action of EM fields, namely $S_{\mathrm{a}}[A]$. Thus, the total partition function $Z_{\textrm{tot}}$ is given as below:
\[ Z_{\textrm{tot}} = \int D[A] \, e^{-S_{\mathrm{a}}[A]} Z[A].      \tag{29} \label{eq29}          \]
Let $A'_{\mu} = (2e)^{-1} \partial_{\mu} \theta_s + A_{\mu}$ be the modified gauge field. Then, the phase $\theta$ can be expressed as $\theta = \theta_v + 2e \int^xdx'_{\mu} A'_{\mu}(x')$. Now, we consider the integration over $\theta_s$. Since the spin wave part $\theta_s$ can be absorbed to $A_{\mu}$ by the gauge transformation at every point $x$ in real space, the integration over $\theta_s$ can be removed from the partition function $Z$ in Eq.~(\ref{eq22}). This procedure is similar to the method in Ref.~\cite{Coleman}, in which the phase term vanishes by only performing the gauge transformation without implementing any integration associated with the fluctuations. However, we assert here that it is necessary to bring out the integration with respect to the gauge fields in order to absorb all possible values of the phase fluctuation. Indeed, if all integral ranges in Eq.~(\ref{eq29}) are infinite, the integration of Eq.~(\ref{eq29}) diverges and needs to be fixed by some regularization scheme. We present a prescription for dealing with this problem in Appendix~\ref{AppendixA}. Under such a procedure, the integrations for the phase fluctuation and gauge fields in Eq.~(\ref{eq29}) become converge, and the total partition function $Z_{\mathrm{tot}}$ is obtained as below:
\begin{align*} 
Z_{\textrm{tot}} & = e^{-F_{\textrm{mf}}(Q)/T }  \\
&\quad \times \!\! \int \!D[\theta_v,A'] \, e^{-S_{\textrm{I}}[\theta_v,A']-S_{\textrm{fluc,eff}}[\theta_v,A'] - S_{\mathrm{a}}[A']},      \tag{30} \label{eq30}          
\end{align*}
where,
\begin{align*}
S_{\textrm{I}}[\theta_v,A'] &= S_{\textrm{I}}\, [\partial\theta_v + 2e A'], \tag{31a} \label{eq31a} \\
S_{\textrm{fluc,eff}}[\theta_v,A'] &= S_{\textrm{fluc,eff}}\, [\partial\theta_v + 2e A'].   \tag{31b} \label{eq31b}
\end{align*}

It should be noted that the above procedure differs from the widely used method described in Ref.~\cite{Altland}. There, the Gaussian integration with respect to the phase fluctuation is performed at a \textit{given} gauge field, and the longitudinal part of the gauge fields is automatically excluded from the action, leaving only the transverse part. See Appendix \ref{AppendixA} for various definitions of the transverse and longitudinal parts. However, the coefficient resulting from the integration neglected in the literature indeed diverges. By performing the CFR, this coefficient becomes similar to the second term on the r.h.s. of Eq.~(\ref{eq24}), but without mass. Thus, the Goldstone mode associated with the spin wave part of the phase fluctuation appears awkwardly in the method used in Ref.~\cite{Altland} as if we were dealing with a superfluid. The mode has the same dispersion relation as the Nambu-Goldstone mode~\cite{Nambu} but has a different origin. The latter arises from the fluctuating longitudinal gauge field if the microscopic Coulomb interaction is not concerned~\cite{Nambu}. However, as discussed in Appendix \ref{AppendixA}, the Goldstone mode expected to appear in the calculation in Ref.~\cite{Altland} is absorbed by the longitudinal gauge field and thus vanishes. In Appendix \ref{AppendixA}, we also demonstrate that our formulation can give a plasmon mode, a massive mode first predicted in Ref.~\cite{Anderson}, in both the Lorenz gauge when the Coulomb interaction is considered and the Coulomb gauge when the electric energy is \textit{not} neglected.

Let us now discuss the gauge-invariant superfluid stiffness, which represents the response to the EM field. As discussed in Appendix~\ref{AppendixA}, it is obtained by integrating out the longitudinal gauge field, resulting in the effective action of the transverse modified gauge fields. The explicit expression is given by Eq.~(\ref{a5}). At the same time, the superfluid stiffness of the vortex part $\theta_v$ in Eq.~(\ref{eq30}) changes to be given by Eq. (\ref{a5}). In this case, however, it is probably necessary to largely modify the BKT theory due to the angular dependence of the expression. We here leave this problem as a future issue. Below, we will neglect the correction to the superfluid stiffness due to the longitudinal gauge field.

The gauge field fluctuation then leaves only the transverse part. However, this part should be neglected when considering a purely 2D superconductor. It is well known that even in the mean-field theory of the transverse gauge field, i.e., Maxwell's equations, the EM fields screen the interaction between the vortices from the quasi-long-range interaction to the short-range one within the Pearl's length $l_p$ \cite{Pearl}. As described in the introduction, superconductivity occurs in purely 2D systems if the system size is small compared to the Pearl's length $l_p$. Hereafter, this condition is assumed, and the transverse gauge field is neglected.

\subsection{Vortex Fluctuation\label{IV B}}

We now consider the last part of the fluctuation action: the vortex part. This part is responsible for the BKT transition and is composed of Eqs.~(\ref{eq31a}) and (\ref{eq31b}). As described in Sec.~\ref{VI}, Eq.~(\ref{eq31a}) cancels out with a supercurrent source action. Thus, only the action Eq.~(\ref{eq31b}) needs to be analyzed. Assuming that $\theta_v=\theta_v(\bm{x})$, from Eqs.~(\ref{eq25}) and (\ref{eq31b}) the effective fluctuation action is
\[         S_{\textrm{fluc,eff}} =    \frac{1}{2T} \int d^2\bm{x} \,  \tilde{J}_{ij} (\partial_{i} \theta_v) (\partial_{j} \theta_v). \tag{32} \label{Sfa}   \]
Following the discussion in the previous section, we have neglected the $A'_{\mu}$ terms. Let us invoke the mean-field theory of $\theta_v$. This gives $\tilde{J}_{ij}\partial_{ij} \theta_v=0$ almost everywhere except at cores of vortices. To proceed further, we rescale the coordinates as $\tilde{x}_1\equiv x_1/\tilde{J}^{1/2}_{11}$ and $\tilde{x}_2\equiv x_2/\tilde{J}^{1/2}_{22}$. The topological condition for a vortex is here given as $\oint d\tilde{\bm{r}}\cdot ( \nabla_{\tilde{\bm{r}}}\theta_v) = 2\pi n_v$, where $n_v\in\mathbb{Z}$, and the subscript of $\nabla_{\tilde{\bm{r}}}$ means that the gradient is performed for $\tilde{\bm{r}} = (\tilde{x}_1, \tilde{x}_2)$. Note that $\tilde{J}_{12}$ is zero in the present model. By this way, the action of the vortex part becomes
\[         S_v =    \frac{1}{2T} \int d^2\tilde{\bm{x}} \,  J_{B} \, (\nabla \theta_v )^2, \tag{33} \label{Svo}   \]
where $J_{B} = \big[ \tilde{J}_{11}\tilde{J}_{22} \big]^{\frac{1}{2}}$ is called the bare effective superfluid stiffness hereafter. For later convenience, we define $K_B \equiv J_{B}/T$.  

If $\tilde{J}_{11}=\tilde{J}_{22}$, vortex cores become circles with diameter $a$. Otherwise, they are ellipses. Let $a_1$ and $a_2$ be the principal diameters of the core. Since the coordinates are being normalized by $\tilde{J}^{1/2}_{ii}$, an elliptical vortex core transforms into a circular one. A straightforward extension from the original KT theory \cite{Kosterlitz1}, in which the vortex core is circular, can be performed by assuming that $a_1 a_2 = a^2$. The vortex core is a circle with diameter $\tilde{a}\equiv a/\sqrt{J_{B }}= \tilde{a}_1=\tilde{a}_2$ in the $\tilde{x}_1\tilde{x}_2$-coordinate system.

\subsection{A Review of BKT Theory\label{IV C}}

Since the effective action has become equivalent to a conventional isotropic action in the $\tilde{x}\tilde{y}$ coordinate system, the remaining procedure can be carried on with the familiar BKT theory. Below, we review the BKT theory in this subsection. Let $\tilde{\bm{r}}^{(i)}$ be the positional vector of a vortex $i$ in the $\tilde{x}\tilde{y}$-coordinate system. Following a standard method, we reach the partition function of the vortex part action $S_v$, namely $Z_v$, as below:
\begin{align*} 
Z_{v} = \sum_{N}^{\infty} & \frac{Y^{2N}_0}{\tilde{a}^{4N}(N!)^2} \int \bigg(\prod_{i=1}^{2N} d^2\tilde{\bm{r}}^{(i)}\bigg) \\
&\times \exp \bigg[ 2\pi^2K_B \sum_{i\not=j} n_{vi} n_{vj} \, C\big(\tilde{\bm{r}}^{(i)} -\tilde{\bm{r}}^{(j)} \big) \bigg],  \tag{34} \label{Zv} 
\end{align*}
where $N$ is the number of vortex-antivortex pairs, $C(\tilde{\bm{r}})=\ln |\tilde{\bm{r}}/\tilde{a}|/(2\pi)$, $Y_0=\exp\lr{-\mu_0/T}$ is the vortex fugacity, and $\mu_0$ is the vortex core energy for the vortex with vorticity $n_v=\pm 1$. For simplicity, we give the form of $\mu_0$ as $\mu_0 = \pi \xi^2 |f_{S}|$ as in Ref.~\cite{Mondal}, where $\xi$ is the BCS coherence length given by $\xi^{-1} = \pi\Delta / v_{\rm F}$~\cite{Mondal, Kramer}, and $f_S$ is the condensation energy, defined in Sec.~\ref{5V A}. This means that the diameter of cores $a$ introduced in the previous section is $a=2\xi$. The picture that circular vortex cores have a radius of order $\xi$ comes from the fact that the scale of the spatial modulation in gap amplitude is an order of $\xi$. This is valid unless the temperature is much lower than the transition temperature where another length scale becomes shorter than $\xi$ \cite{Kramer}. 

\begin{figure}[b]
\centering
\includegraphics[width=5.0cm]{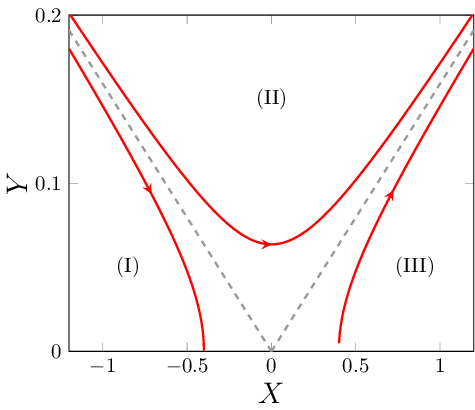}
\caption{The Kosterlitz renormalization group flow with integral constant $C_0 = 4\pi^2Y^2-X^2$. The gray dashed lines, given by $2\pi Y=\pm X$, divide the flows into three regions, I, II, and III. The red solid curves illustrate the flows in each region. The quasi-long-range order occurs only in Region I.}

\label{fig2}
\end{figure}

Following the original KT theory \cite{Kosterlitz1}, we can study the partition function $Z_v$ by transforming the many-body problem into a one-body problem with renormalized coupling constants. There, the interaction between the vortex and antivortex is shielded and normalized by the dielectric coefficient $\eps(\tilde{r})$ at a distance $\tilde{r}$. With this, the action of vortex-antivortex pairs per unit area of films, namely $S_{vp}$, becomes
\begin{align*}
-\frac{1}{A}\log Z_{v} &\simeq S_{vp} \\
&\equiv -\frac{2\pi Y^2_0}{a^2} \int_{1}^{\infty} du\, u  \exp \Big[   -\frac{4\pi^2 K_B C(\tilde{a}u)}{\eps(\tilde{a} u)}    \Big].  \tag{35}  \label{fvo}  \end{align*}
Let us define $K(\tilde{r})\equiv K_B/\tilde{\eps}(\tilde{r})$ and rewrite it as a function of the scale factor $l\equiv \ln |\tilde{r}/\tilde{a}|$, i.e., $K(\tilde{r}) = K(l)$. In the same way, we define the renormalized superfluid stiffness at scale $l$, namely $J(l)$, as $J(l)\equiv TK(l)$ for later convenience. The $K(l)$ is expressed as follows:
\[ [K(l)]^{-1} = K^{-1}_B + 4\pi^3 \int_0^{l} dl'\, [Y(l')]^2, \tag{36a}  \label{Kl}  \]
with
\[Y(l) = Y_0 \exp\Big[ 2l - \pi \int_0^l dl'\,  K(l')\Big].  \tag{36b} \label{yl} \]
In a sophisticated method of real-space renormalization group \cite{Kosterlitz2}, it is shown that $K(l)$ and $Y(l)$ obey the following Kosterlitz renormalization group equations:
\[\D{}{l}K^{-1} = 4\pi^3Y^2 + O(Y^4), \tag{37a} \label{RGK}\]
\[\D{}{l}Y = (2-\pi K) Y + O(Y^2). \tag{37b} \label{RGy}\]
By neglecting the higher-order terms in these equations, Eqs.~(\ref{Kl}) and (\ref{yl}) are derived. Let us introduce $X(l) = 1 - \frac{\pi}{2} K(l)$ and substitute it into the above equations. Note that these parameters $X(l)$ and $Y(l)$ are not the spatial coordinates. We will write them as $X$ and $Y$ in some places in this section for brevity. Neglecting $O(XY^2)$,~$O(Y^4)$, $O(X^2Y)$, and $O(Y^3)$ terms, we obtain the differential equations, $dX/dl = 8\pi^2Y^2$ and $dY/dl = 2XY$. From these equations, we have 
\[ 4\pi^2[Y(l)]^2 - [X(l)]^2 = C_0, \tag{38} \label{BKTinr} \]
where $C_0$ is a real number. It should be noticed that renormalization group flows of $X(l)$ and $Y(l)$ strongly depend on the signs of $C_0$ and $X(l)$. The solutions are classified into three regions by the lines $2\pi Y(l) = \pm X(l)$, as depicted in Fig.~\ref{fig2}. The renormalization group flow in each region does not cross over to another region. First, the region with $C_0<0$ and $X(l)<0$ is termed as \textit{Region I}. In this region, $X(l)$ and $Y(l)$ are given as follows:
\begin{align*}
X(l) &= -\sqrt{|C_0|} \, \frac{1-D_0\,e^{-4l\sqrt{|C_0|} } }{1+D_0\, e^{-4l\sqrt{|C_0|}} },  \tag{39a} \label{XX1} \\
Y(l) &= \frac{1}{\pi}\sqrt{|C_0D_0|} \, \frac{ e^{-2l\sqrt{|C_0|} } }{1+D_0\, e^{-4l\sqrt{|C_0|} } },  \tag{39b} \label{yl1} 
\end{align*}
where $D_0= \frac{\sqrt{|C_0|}+X(0)}{\sqrt{|C_0|}-X(0)}$ and $X(0)=1- \frac{\pi}{2} K_B$. In Region I, all renormalization group flows end at $Y=0$ and $X<0$, meaning that the renormalized superfluid stiffness at scale $l$, $K(l)$, is still finite even if $l$ is pushed to infinity. Second, the region with $C_0>0$ is termed as \textit{Region II}, and $X(l)$ and $Y(l)$ are obtained as
\begin{align*}
X(l) &= ~\sqrt{C_0} \tan  \Big( \alpha_0 + 2l\sqrt{C_0} \Big), \tag{40a} \label{XXX2} \\
Y(l) &= \frac{1}{2\pi}\sqrt{C_0} \sec  \Big( \alpha_0 + 2l\sqrt{C_0} \Big), \tag{40b} \label{yl2}
\end{align*}
where $\alpha_0 = \frac{1}{\sqrt{C_0}}\arctan \big( X(0)/\sqrt{C_0} \big) $. In this region, $K(l)$ becomes zero at various values of $l$ with $Y(l)$. However, due to the trigonometric functions of Eqs.~(\ref{XXX2}) and (\ref{yl2}), the physical solutions, in which $K(l)$ is non-negative, are restricted to some regions. With this consideration, the scale factor $l$ in Region II is limited by the value at which $X(l)$ becomes zero for the first time as the scale factor $l$ increases. Thus, all renormalization group flows end at points with $K(l)=0$. Lastly, the region with $C_0<0$ and $X(l)>0$ is termed as \textit{Region III}. The renormalization group flows in this region end up with $K(l)=0$, the same as Region II. The renormalization group flows in the three regions are illustrated in Fig.~\ref{fig2}.

As discussed above, the renormalized superfluid stiffness at the terminate point of a renormalization group flow is finite in Region I while zero in Region II and III. Hereafter, the superfluid stiffness at the terminate point is called the full renormalized superfluid stiffness $J_R$, and we denote $K_R\equiv J_R/T$. In particular, $K_R$ is given as $K_R=K(l\rightarrow\infty)$ in Region I.

Now, let us discuss the correlations of vortices. It is known that the interaction between vortices becomes a quasi-long-range interaction in Region I, while it is short-range in Region II and III. Therefore, there is a phase transition between Region I and the other regions. This transition is well known as the BKT transition. At the BKT transition temperature, namely $T_{\rm c}$, the system is located on the boundary line $2\pi Y(l) = -X(l)$. There, the renormalization group flow terminates at $Y=X=0$ if $l$ goes to infinity. In this way, we obtain a universal relation at the BKT transition as follows:
\[    \pi J_R/T_{\rm c} =    \pi K_R = 2.     \tag{41} \label{BKT}                       \]
This relation is called the NK relation~\cite{Nelson} in the superfluid systems. In this paper, the NK relation for the full renormalized superfluid stiffness is simply termed the NK criterion. This criterion is widely used in purely 2D superconductors \cite{Larkvv}.

\section{Phase diagram and superfluid response\label{5V}}

In this section, we study the phase diagram of the model at zero current and the superfluid stiffness at zero and finite current. We first describe the equations used in the numerical calculations and then discuss the results in detail.

\begin{figure*}[t]
\begin{minipage}{0.32\hsize}
\subfloat[\label{fig3a}]{\includegraphics[width=5.9cm]{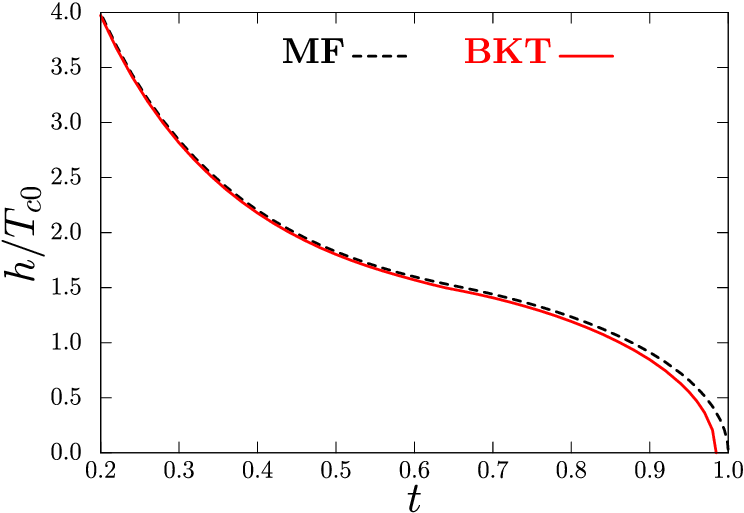}}
\end{minipage}
\begin{minipage}{0.32\hsize}
\subfloat[\label{fig3b}]{\includegraphics[width=5.9cm]{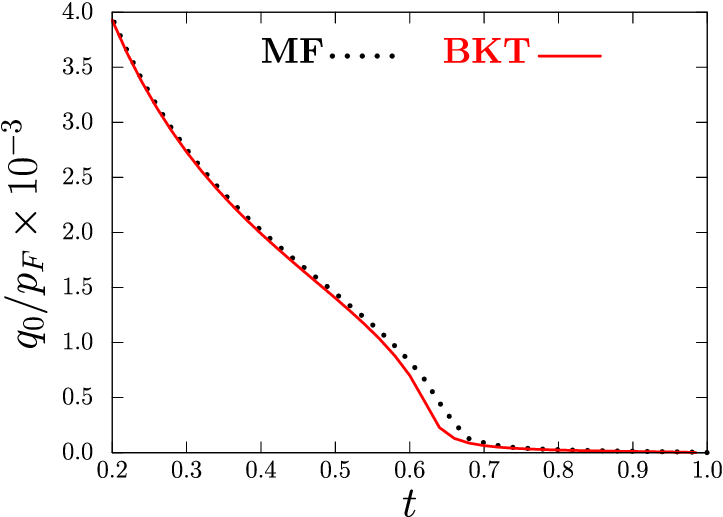}}
\end{minipage}
\begin{minipage}{0.32\hsize}
\subfloat[\label{fig3c}]{\includegraphics[width=5.9cm]{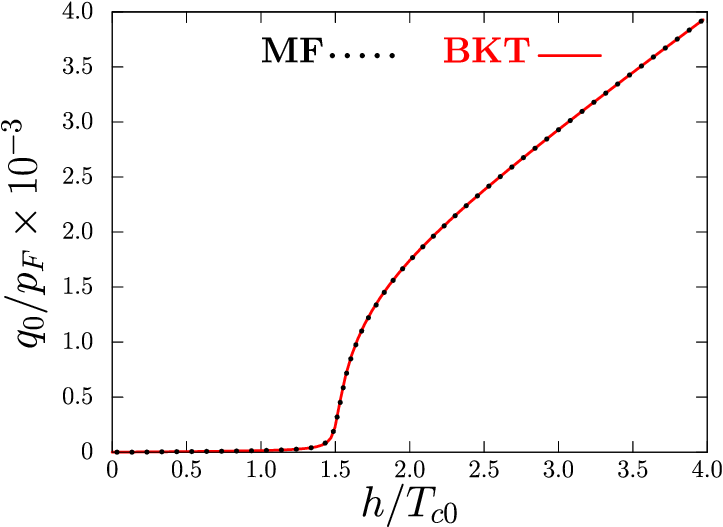}}
\end{minipage}

\caption{(a) The $h$-$t$ phase diagram. The black dashed line is the mean-field transition line, while the red solid line shows the BKT transition line. (b) and (c) show the Cooper pair momentum $q_0$ as a function of the reduced temperature $t$ and magnetic field $h$, respectively. The red (black dotted) lines represent the $q_0$ on the BKT (mean-field) transition line in the figure (a). }

\label{fig3}
\end{figure*}

\subsection{Equations \label{5V A}}

To obtain any result from now on, it is necessary to clarify how to calculate the mean-field free energy. Here, we use the same procedure as the BCS theory and introduce an energy cutoff with width $2\Lambda$, restraining the momentum integration in Eqs.~(\ref{eq10}) and (\ref{eq11}) to the interval from the lower bound $p_- = p_{\rm F} \sqrt{1-\Lambda/E_{\rm F}}$ to the upper bound $p_+ = p_{\rm F}\sqrt{1+\Lambda/E_{\rm F}} $. Strictly speaking, the cutoff momentum $p_{\pm}$ depends on $\bm{Q}$, making the analysis harder. However, as long as $|\bm{Q}|/p_{\rm F} \ll \Lambda/E_{\rm F}$, such dependence is negligible. Taking this procedure, we solved the gap equation
\[      \pd{}{\Delta} F_{\mathrm{mf}}[\Delta;\bm{Q}] = 0,       \tag{42} \label{Ga}  \]
to obtain a gap $\Delta$ for a Cooper pair momentum $\bm{Q}$. Since the parity symmetry is broken along the $x_2$-direction at finite magnetic fields, only the states with finite $Q_2$ can give the most stable states of the superconductor. In the following, in addition to $Q_0$, the limit $Q_1\rightarrow 0$ is also taken after some successful calculations, and we denote $q_0$ as the Cooper pair momentum $Q_2$ of the state with the lowest free energy at a temperature $T$ and a magnetic field $h$. The reduced temperature is introduced as $t=T/T_{\rm c0}$, where $T_{\rm c0}$ is the mean-field transition temperature at zero magnetic field. In the numerical computation, we set $\Lambda =0.1324 \, E_{\rm F}$, $E_{\rm F}/T_{\rm c0} = 10^3$, $gN(0) = 0.2$, and $\alpha =0.01 v_{\rm F}$. Here, $N(0)=mA_{\mathrm{s}}/(2\pi)$ is the density of states at the Fermi level.

It has to be noted that the mean-field free energy density $f_{\mathrm{mf}}(\bm{Q})$ at a given $\bm{Q}$ is composed of the superconducting part and the normal part: the latter comes from the momentum space in the interval $[0,p_-]$. Here, $Q_0=0$ has already been substituted, and the spatial components of the supercurrent and the total superfluid stiffness are needed for numerical calculations. We can regulate $f_{\mathrm{mf}}(\bm{Q})$ by subtracting the normal state free energy density $f_{\rm N}= f_{\mathrm{mf}}[0;\bm{Q}]= f_{\mathrm{mf}}[0;0]$, in which all momenta are included. By defining the condensation energy density $f_{\rm S}(\bm{Q})$ as $f_{\rm S}(\bm{Q}) \equiv f_{\mathrm{mf}}(\bm{Q}) - f_{\rm N}$, the SC states at zero current are then given by minimizing this $f_{\rm S}(\bm{Q})$. We can also replace $ f_{\mathrm{mf}}(\bm{Q})$ by $f_{\rm S}(\bm{Q})$ in Eqs.~(\ref{eq28}) to calculate any superfluid stiffness.

We obtain the mean-fiend transition line by first taking the limit of Eq.~(\ref{Ga}) when $\Delta$ approaches zero. This is the linearized gap equation, whose solutions give the critical magnetic field $h$ for a given $Q_2$ at each temperature. Finding $Q_2$ that maximizes $h$, we eventually obtain the mean-field transition line.

To find the BKT transition temperature at a magnetic field $h$, namely $T_{\rm c}(h)=t_{\rm c}(h)T_{c0}$, we solve the NK criterion of Eq.~(\ref{BKT}) that corresponds to the relation of bare parameters $2\pi Y(0) = -X(0)$ (see Sec.~\ref{IV C}). We emphasize that the parameters $\Delta,~J_B$, and $f_S$ depend on the temperature $t$, the magnetic field $h$, and the Cooper pair momentum $Q_2$. Below, we refer to $T_{\rm c} \equiv T_{\rm c}(0)$ as the zero field BKT transition temperature. Moreover, the bare effective superfluid stiffness $J_B$ used in calculating the BKT transition temperature at each magnetic field is calculated with $Q_2$ at which the stable mean-field solution gains the maximum condensation energy.

\subsection{Phase Diagram\label{5V B}}

We show the $h$-$t$ phase diagram and the temperature and magnetic field dependences of $q_0$ along the mean-field and BKT transition lines in Figs.~\ref{fig3a} \ref{fig3b}, and \ref{fig3c}. The mean-field transition line looks to reproduce the clean limit of a similar model~\cite{Ilic}, indicating the correctness of numerical calculations. We have also confirmed that just below the mean-field transition line, the magnitude relation between the critical currents in the forward and backward directions along the $x_2$ axis does not change, consistent with Ref.~\cite{Ilic}. The BKT transition temperature lies slightly lower than the mean-field transition temperature [Fig.~\ref{fig3a}]. The $q_0$ is also smaller at the BKT transition line than at the mean-field line. The $q_0$-$h$ lines, however, almost coincide with each other [Fig.~\ref{fig3c}]. Detailed analysis of the region near the BKT transition temperatures around $t=0.6$ shows that at a fixed magnetic field $q_0$ slightly increases as the temperature decreases, and thus the $q_0$-$h$ line at the mean-field transition and that at the BKT transition do not exactly overlap. Nevertheless, our results reflect a weak dependence of $q_0$ on temperature. Since the $h$ dependence of $q_0$ is significant, we can see a clear deviation in the $q_0$-$t$ lines [Fig.~\ref{fig3b}].

\begin{figure}[b]
\captionsetup{font=normal,justification=raggedright,singlelinecheck=false}
\begin{minipage}{0.49\hsize}
\subfloat[$h=0$\label{figx1}]{\includegraphics[width=4.2cm]{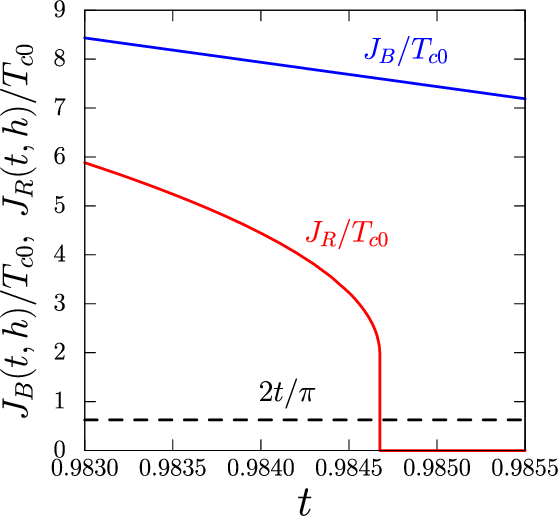}}
\end{minipage}
\begin{minipage}{0.49\hsize}
\subfloat[$h=1.57T_{\rm c0}$\label{figx2}]{\includegraphics[width=4.2cm]{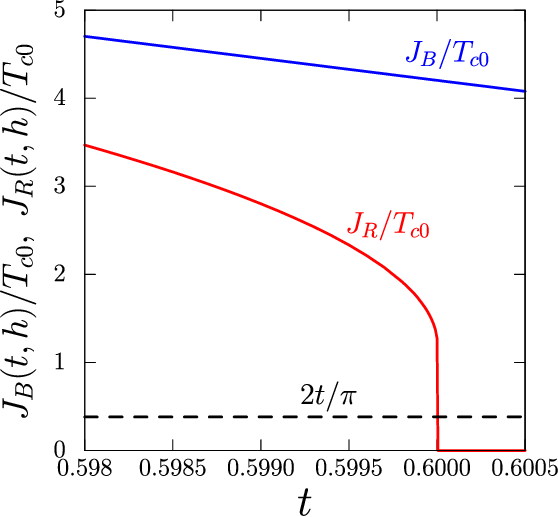}}
\end{minipage}

\caption{The temperature dependence of the bare effective superfluid stiffness $J_B(t,h)$ and full renormalized superfluid stiffness $J_R(t,h)$ at fixed magnetic fields (a) $h=0$ and (b) $h=1.57 T_{\rm c0}$. The blue and red lines show $J_B(t,h)/T_{\rm c0}$ and $J_R(t,h)/T_{\rm c0}$, respectively. The dashed black lines show $2t/\pi$ for the NK criterion.}

\label{fig4}
\end{figure}

Readers may be interested in the temperature dependence of the superfluid stiffness when the Cooper pair momentum becomes finite in the helical SC state at finite magnetic fields. However, as shown in Fig.~\ref{fig4}, the temperature dependence of the bare effective superfluid stiffness $J_B$ and the full renormalized superfluid stiffness $J_R$ in a finite magnetic field is qualitatively the same as those at zero magnetic field. This implies a weak dependence of the superfluid stiffness on the Cooper pair momentum $q_0$ near the BKT transition line.

Here we give a comment on the phase diagram in the low-temperature region. In Fig.~\ref{fig3a}, we show the BKT and mean-field transition lines because the aim of this study is to clarify nonlinear transport phenomena near the BKT transition line. When we are interested in the low-temperature region, we should consider the stripe phase which was shown to be stable in the mean-field theory~\cite{Agterberg, Aoyama}. According to the renormalization group analysis for centrosymmetric systems~\cite{Berg}, the stripe phase does not melt to the helical phase but can become the so-called charge $4e$ superconducting phase, i.e., a phase without spatial modulation due to Cooper pairs but with fractional quantum vortices. Similar phase may appear in noncentrosymmetric superconductors at low temperatures because the stripe phase is stable there in the mean-field theory. It is desirable to clarify the low-temperature phase diagram of purely 2D noncentrosymmetric superconductors.

Let us then discuss the BKT transition temperature at zero magnetic field, which is obtained as $T_{\rm c}=0.98467(5) \, T_{\rm c0}$ in our parameters. Although the width of the critical region defined by $ \tau_{\rm c} \equiv 1-T_{\rm c}/T_{\rm c0}$ is small $\tau_{\rm c} \approx 0.0153$, the deviation of the mean-field transition magnetic field $h_{\mathrm{mf}}(T)$ from the BKT transition magnetic field $h_{\rm c}(T)$, namely 
\[\Delta h(T) \equiv h_{\mathrm{mf}}(T) - h_{\rm c}(T),    \tag{43} \label{hd}         \]
is large compared to the 2D Pauli limiting field $h_{\rm p}/T_{\rm c0}=1.24$ \cite{Clogston}. For example, we obtain $\Delta h(T_{\rm c})/h_{\rm p} \approx 0.30$.

\begin{figure}[tbp]
\begin{minipage}{\hsize}
\subfloat[\label{fig5a}]{\includegraphics[width=6.0cm]{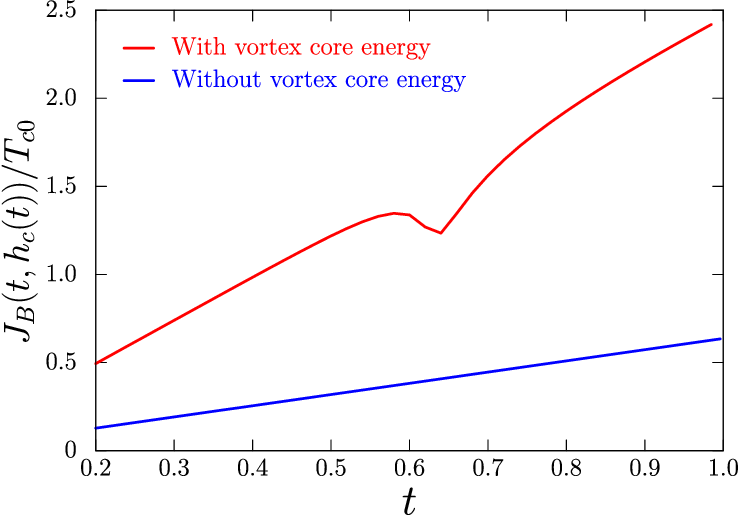}}
\end{minipage}
\begin{minipage}{\hsize}
\subfloat[\label{fig5b}]{\includegraphics[width=6.0cm]{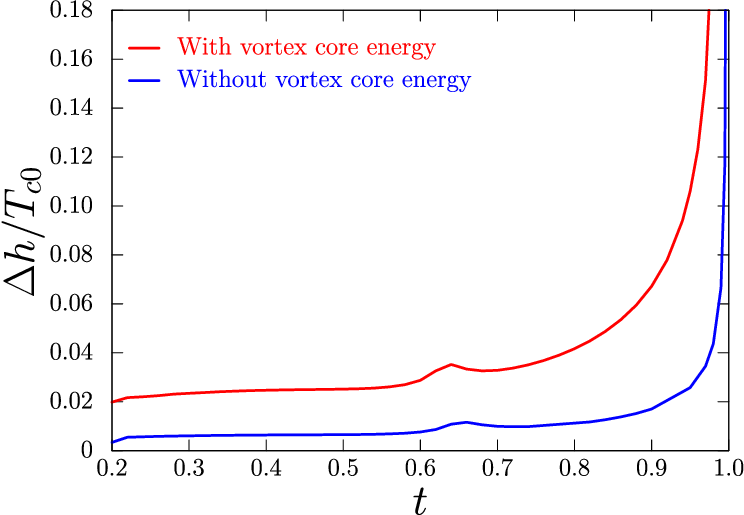}}
\end{minipage}
\caption{The temperature dependence of (a) the bare effective superfluid stiffness $J_B(t,h_c(t))$ and (b) the difference in critical magnetic fields between the mean-field theory and the BKT theory $\Delta h$. The results along the BKT transition line are plotted. The red and blue lines are obtained with and without considering the vortex core energy. }

\label{fig5}
\end{figure}

It is noted that the BKT transition line in Fig.~\ref{fig3a} takes into account the vortex core energy. In many pieces of literature, the BKT transition temperature is calculated by neglecting the vortex core energy. Such an approximation can lead to two effects on the results. One is that the BKT transition temperature at zero magnetic field is overestimated. When the vortex core energy is neglected, the zero field BKT transition temperature is obtained as $T'_{\rm c} = 0.99592(7) \, T_{\rm c0}$. This value is consistent with the formula $T'_{\rm c}\simeq T_{\rm c0}(1-4Gi)$, where $Gi \simeq T_{\rm c0}/E_{\rm F}$ is the Ginzburg-Levanyuk number for 2D clean superconductors~\cite{Larkvv}. Correspondingly, $\Delta h(T)$ defined in Eq.~\eqref{hd} is decreased as $\Delta h(T'_c)/h_{\rm p} \approx 0.15$ at $T'_c$, which is about half of the result including the vortex core energy. The other effect is seen in the blue line of Fig.~\ref{fig5a}: the bare effective superfluid stiffness along the BKT transition line becomes linear with temperature. This feature is trivial because the NK criterion is given by $T'_c \approx \pi J_B/2$ in the formulation neglecting the vortex core energy. However, it is shown by the red line in Fig.~\ref{fig5a} that the bare effective superfluid stiffness is nonlinear and non-monotonic with temperature due to the vortex core energy.

Finally, we discuss $\Delta h(T)$ in Eq.~(\ref{hd}) below the zero-field BKT transition temperature. As shown in Fig.~\ref{fig5b}, $\Delta h(T)$ (red line) is not monotonically decreasing upon cooling the temperature but has a local maximum at $t\approx0.64$. Since the blue line obtained by neglecting the vortex core energy exhibits qualitatively the same behavior, the non-monotonic behavior should stem from the dependence of the bare effective superfluid stiffness on temperature and magnetic field, but not from the vortex core energy. Therefore, the origin of the non-monotonic behavior in $\Delta h(T)$ has to be distinguished from that of the non-monotonic bare effective superfluid stiffness in Fig.~\ref{fig5a}. However, due to the temperature dependence of the vortex core energy, the local maximum of $\Delta h(T)$ appears at a slightly lower temperature than that obtained without vortex core energy.

\subsection{Condensation Energy and Superfluid Stiffness around helical crossover\label{5V C}}

\begin{figure}[b]
\captionsetup{font=normal,justification=raggedright,singlelinecheck=false}
\begin{minipage}{0.49\hsize}
\subfloat[$h_c(t=0.7)$\label{fig6a}]{\includegraphics[width=4.2cm]{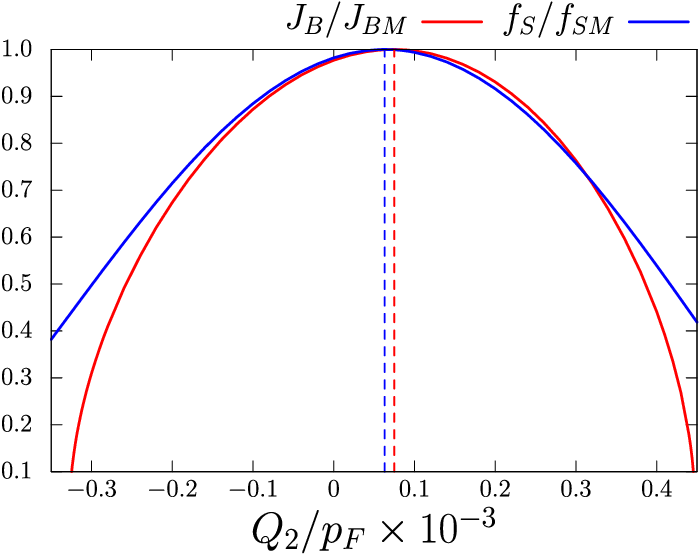}}
\end{minipage}
\begin{minipage}{0.49\hsize}
\subfloat[$h_c(t=0.64)$\label{fig6b}]{\includegraphics[width=4.2cm]{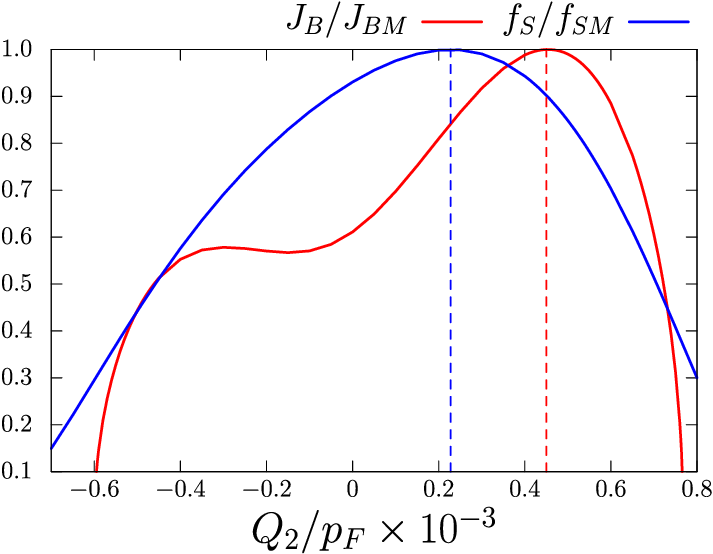}}
\end{minipage}
\begin{minipage}{0.49\hsize}
\subfloat[$h_c(t=0.6)$\label{fig6c}]{\includegraphics[width=4.2cm]{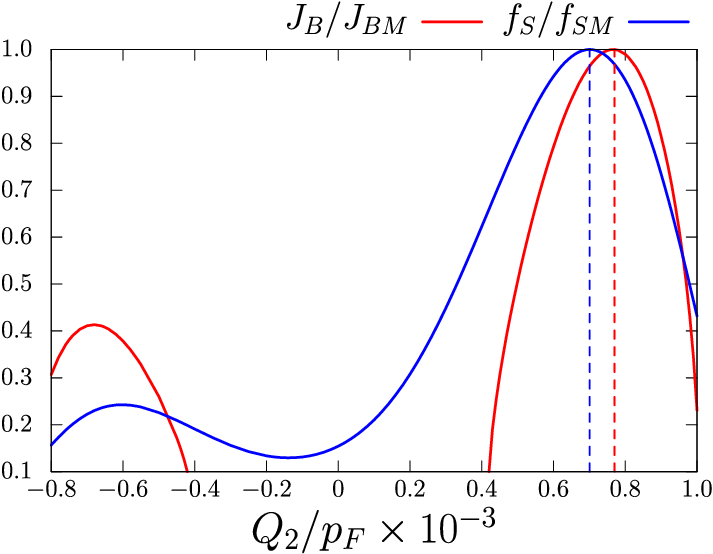}}
\end{minipage}
\begin{minipage}{0.49\hsize}
\subfloat[$h_c(t=0.5)$\label{fig6d}]{\includegraphics[width=4.2cm]{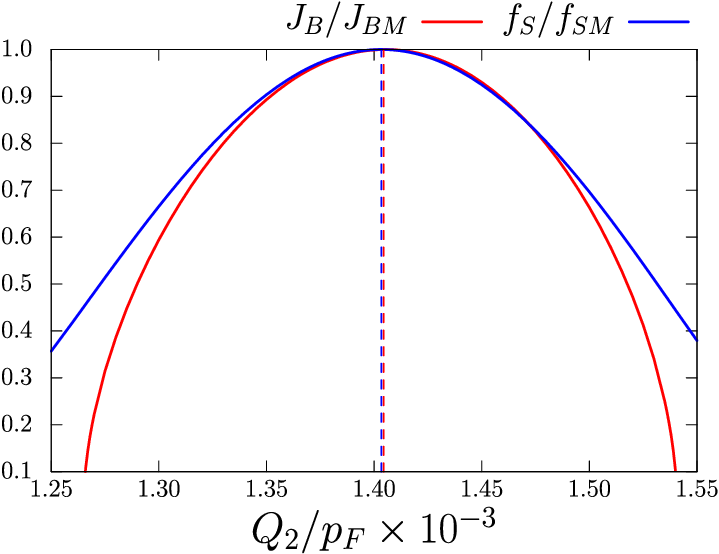}}
\end{minipage}
\captionsetup{font=normal,justification=raggedright,singlelinecheck=false}

\caption{The Cooper pair momentum $Q_2$ dependence of the bare effective superfluid stiffness $J_B>0$ and condensation energy $f_S<0$ at various temperatures on the BKT transition line. We plot the quantities $J_B$ and $f_S$ scaled by their global maximum and minimum, $J_{BM}$ and $f_{SM}$, respectively. The blue and red dashed lines indicate $q_0$ and $q_{J}$, respectively. Around the temperature and magnetic field in panel (b), the Cooper pair momentum $q_0$ of the stable SC state drastically increases, as seen in Fig.~\ref{fig3b}. Thus, the parameters are located near the crossover line of helical SC states. Panel (a) shows the small-$q_0$ helical SC state, while panels (c) and (d) show the large-$q_0$ helical SC state. In panel (d), we do not show the metastable state, which appears at $Q_2<0$.} 

\label{fig6}
\end{figure}

The non-monotonic behavior of $\Delta h(T)$ makes us curious about the superconducting properties around the local maximum of $\Delta h(T)$. Indeed, we find two characteristic features in the condensation energy and superfluid stiffness.

First, $q_J$ denoting the momentum at which the bare effective superfluid stiffness reaches its maximum significantly deviates from $q_0$. Although $q_J$ almost coincides with $q_0$ at most temperatures and magnetic fields, these two momenta are considerably different around the magnetic field corresponding to the local maximum of $\Delta h(T)$, as shown in Figs.~\ref{fig6a}-\ref{fig6d}. The difference between $q_0$ and $q_J$ can be quantitatively evaluated by the GL theory. Let us consider the GL free energy
\[   f_{\mathrm{GL}} = a\Delta^2 + \frac{1}{2}b \Delta^4,     \tag{44} \label{Gl}               \]
where $a$ and $b$ are functions of $T$, $h$, and $\bm{Q}$. Optimizing $f_{\mathrm{GL}}$, we have $\Delta^2 = -a/b$ and $f_{\mathrm{GL}} = -a^2/2b$. When we neglect the momentum dependence of $b$ for simplicity, from Eq.~(\ref{eq28}), we have 
\[\tilde{J}_{22} = -(a'^2 +  aa'')/b,  \tag{45} \label{GLj}  \] 
where the prime means the derivative with respect to $Q_2$. The saddle point condition for the above equation is $3a'a''+aa'''=0$. When $f_{\mathrm{GL}}$ is minimized with respect to $Q_2$, $a'=0$ is satisfied. Thus, $\tilde{J}_{22}$ reaches its maximum at the momentum different from $q_0$ in general. To clarify the condition, let us expand $a$ around the momentum $Q_2=q_0$ where $a'=0$ is satisfied. We then have $a \simeq a_0 \, (1+a_2 \delta Q_2^2 +a_3\delta Q_2^3 +a_4\delta Q_2^4)$, where $\delta Q_2 = Q_2 - q_0$ with $a_i$ being a function of $T$, $h$, and $Q^2_1$. With this expansion, we find that $\tilde{J}_{22}$ reaches its maximum at different momentum from $q_0$ if $a_3\not=0$. It can be shown in the same way that $\tilde{J}_{11}$ is also maximized at the momentum away from $q_0$ in general. Thus, $q_J$ is different from $q_0$. This conclusion does not change when the momentum dependence of $b$ is included, although the analysis becomes complicated.

Taking the momentum dependence of $b$ into account, we also show that the momentum at which the gap amplitude reaches its maximum, namely $q_{\Delta}$, can be different from $q_0$ and $q_J$. From Eq.~(\ref{Gl}), the saddle condition for $\Delta$ is $(a'b - ab')/b^2=0$, while that for $f_{\mathrm{GL}}$ is $a \, (2a'b-ab')/b^2=0$. Therefore, the momentum dependence of $b$ leads to the difference between $q_{\Delta}$ and $q_0$. We can also show that $q_{\Delta}$ is not equal to $q_J$ when $a$ and $a'''$ are finite. The Cooper pair momentum different from $q_0$ in the equilibrium state can be realized under the supercurrent. The superconducting state carrying the supercurrent will be discussed in Sec. \ref{VI}.

Second, a metastable state appears as shown in Fig.~\ref{fig6c}. At low magnetic fields, the magnitude of condensation energy shows a single peak as a function of $Q_2$ [Fig.~\ref{fig6a}]. However, when the magnetic field is increased along the BKT transition line, we see the two-peak structure as in Fig.~\ref{fig6c}, which indicates the presence of a metastable SC state. This is essentially the same behavior as shown in Ref.~\cite{Daido1} and associated with the helical crossover where the Cooper pair momentum $q_0$ drastically increases. The metastable states appear in the magnetic field higher than the crossover line.

The two features discussed above are closely related to each other. From the above analysis of the GL free energy, we notice that the cubic term with the coefficient $a_3$ is essential for deviation of $q_J$ from $q_0$. Interestingly, this term is also essential for the intrinsic SD effect~\cite{Daido1,Yuan,He-Nagaosa,Daido2} and nonreciprocal paraconductivity and Hall effect~\cite{Daido_para}. It has been shown that the cubic term is enhanced around the crossover line of helical states~\cite{Daido1,Yuan,Daido_para}. Thus, the separation between $q_J$ and $q_0$ should be enhanced around the crossover line. Our calculation results of Figs.~\ref{fig6b}-\ref{fig6d} support this expectation, as the separation between $q_J$ and $q_0$ is large near the crossover line. Therefore, the appearance of the metastable state and the pronounced difference between $q_J$ and $q_0$ have the same origin, the crossover in helical SC states. Our results have also shown that the crossover of helical SC states gives rise to characteristic behaviors of the BKT transition. The non-monotonic behavior of $\Delta h$ originates from the crossover phenomena: $\Delta h$, defined from Eq.~(\ref{hd}) shows the peak in both calculations including and excluding the vortex core energy [Fig.~\ref{fig5b}]. 

\subsection{Anisotropy of Superfluid Stiffness \label{5V D}}

\begin{figure}[t]
\captionsetup{font=normal,justification=raggedright,singlelinecheck=false}

\includegraphics[width=6.0cm]{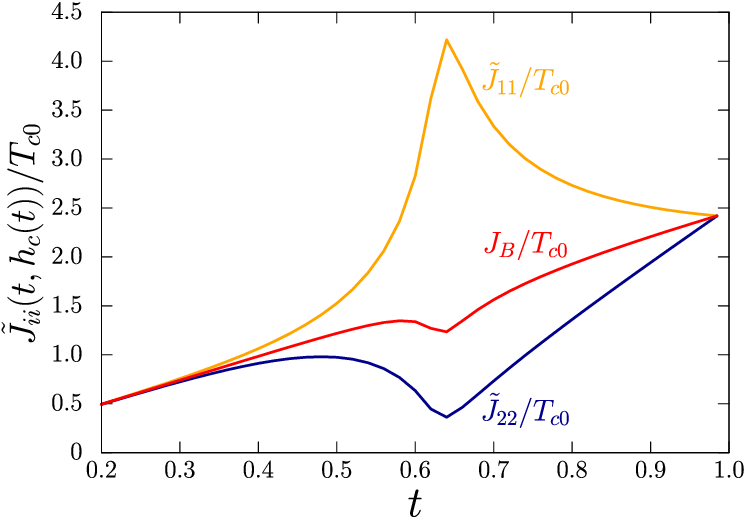}
\caption{The temperature dependence of the total superfluid stiffness along the $x_1$-axis $\tilde{J}_{11}(t,h_c(t))$ (orange line) and the $x_2$-axis $\tilde{J}_{22}(t,h_c(t))$ (blue line). The red line shows the bare effective superfluid stiffness $J_B = \sqrt{\tilde{J}_{11} \tilde{J}_{22}}$.}

\label{fig7}
\end{figure}

So far, we have tackled the effective superfluid stiffness $J_{B}$ obtained by scale transformation. At the end of this section, let us look at the behavior of the total superfluid stiffness, namely, $\tilde{J}_{11}$ and $\tilde{J}_{22}$. As shown in Fig.~\ref{fig7}, the $x_1$-component $\tilde{J}_{11}$ and the $x_2$-component $\tilde{J}_{22}$ show qualitatively different behaviors as decreasing the temperature along the BKT transition line. Both $\tilde{J}_{11}$ and $\tilde{J}_{22}$ show extremum at the temperature $t\approx0.64$ corresponding to the crossover in the helical SC states. The $\tilde{J}_{11}$ perpendicular to the Cooper pair momentum $q_0$ shows the maximum, while the $\tilde{J}_{22}$ parallel to $q_0$ shows the minimum. The non-monotonic and contrasting behaviors of the superfluid stiffness $\tilde{J}_{11}$ and $\tilde{J}_{22}$ are characteristic of the helical superconductors, and signature of helical SC state more drastically appears than in the bare effective superfluid stiffness $J_{B} = \big[ \tilde{J}_{11}\tilde{J}_{22} \big]^{\frac{1}{2}}$. Interestingly, the in-plane anisotropy is remarkably enhanced in the crossover region, and it can be verified by experiments. We note that the behavior of the $x_1$-component is different from the result in the 2D topological FFLO neutral Fermi gas~\cite{Xu2}, where $\tilde{J}_{11}$ decreases with decreasing the temperature along the BKT transition line. Thus, a part of the results is expected to depend on the model.

\section{Supercurrent\label{VI}}

In the following sections, we discuss the superconducting properties under the supercurrent. 
In this section, we formulate the supercurrent. As is well known, Maxwell's equations give a general expression linking an electric current to an action. The SC system is dissipationless even when a supercurrent flows, and the free energy should be defined as long as the superconductivity is not broken. 
Therefore, we expect that a non-dissipative SC state carrying the supercurrent can be determined by extremizing a thermodynamic function. For a while, let us limit ourselves to a mean-field 2D superconductor connected to a current source that controls the electric current. Then, the critical phenomena and EM responses are treated in the same way as in a 3D bulk superconductor. A treatment of a purely 2D superconductor is explained later in this section. 

Let us assume a stationary supercurrent density $2e \bm{j}_s$ flowing through the system. At the mean-field level, the SC state carrying a given supercurrent density $2e\bm{j}_s$ should be determined by minimizing the functional
\[  G[\bm{j}_s,\tilde{\bm{A}}] =  F_{\textrm{mf}}[\tilde{\bm{A}}] - 2e\int d^2\bm{x}\, \bm{j}_{s}(\bm{x}) \cdot \tilde{\bm{A}}(\bm{x}), \tag{46}  \label{GF}  \]
with respect to $\tilde{\bm{A}}$. Here, $\tilde{A}_{\mu}$ termed the total gauge field is given by $\tilde{A}_{\mu} = (2e)^{-1} \partial_{\mu} \Theta +\mathcal{A}_{\mu}$, where $\mathcal{A}_{\mu}$ is a gauge field and $\Theta$ is the total phase of the superconducting order parameter. These quantities include not only the fluctuation part but also the mean-field part, and thus differ from $A_{\mu}$ and $\theta$ introduced in Sec.~\ref{IV} for fluctuation components. The condition $\tilde{A}_0=0$ is assumed in order to keep the thermal equilibrium. $F_{\textrm{mf}}[\tilde{\bm{A}}]$ denotes the mean-field free energy, and its dependence on $\tilde{\bm{A}}$ comes from a straightforward generalization: The effective action in Eq.~(\ref{eq4}) should be expanded around $\tilde{\bm{A}}$, not just the Cooper pair momentum $\bm{Q}$, in general. A similar form of the variational principle was proposed in Ref.~\cite{Samokhin}. The extrema of $G[\bm{j}_s,\tilde{\bm{A}}]$ can be found by solving the saddle point equation $\delta G[\bm{j}_s,\tilde{\bm{A}}]/\delta\tilde{A}_{i}(\bm{x}) = 0$. Let us assume that $G[\bm{j}_s,\tilde{\bm{A}}]$ reaches the global minimum at $\tilde{\bm{A}} = \tilde{\bm{A}}_{m}$, where the functional $G[\bm{j}_s, \tilde{\bm{A}}]$ is reduced to a Gibbs free energy $G[\bm{j}_s] \equiv G[\bm{j}_s,\tilde{\bm{A}}_{m}]$, and the equation for the supercurrent density
\[    j_{si}(\bm{x}) = \frac{1}{2e}\bigg( \frac{\delta}{\delta \tilde{A}_i(\bm{x})} F_{\mathrm{mf}}[\tilde{\bm{A}}] \bigg)_{\tilde{\bm{A}}=\tilde{\bm{A}}_m},      \tag{47}   \label{aj}    \]
is held. Except for the condition $\tilde{\bm{A}} =\tilde{\bm{A}}_m$, this equation is the same as the expression in Refs.~\cite{Daido1, Daido2}. By this variational principle, we can uniquely determine the superconducting states from the metastable states that appear in the mean-field theory of helical superconductors~\cite{Nunchot}.

The variational principle stated above corresponds to the replacement of the pure EM action $S_{\mathrm{a}}$ in Eq.~(\ref{eq29}) by a supercurrent source action $S_{\mathrm{J}}$ given as
\[   S_{\mathrm{J}}[\tilde{A}]    = -2e\int d^3x\, j_{s\mu}(x)\tilde{A}_{\mu}(x).   \tag{48}  \]
The replacement is allowed if the extrema of Eq.~(\ref{GF}) satisfies the Maxwell's equations. Note that $\tilde{A}_{\mu}$ includes the mean-field part $\tilde{A}_{m\mu }$ and the fluctuation part $\delta \tilde{A}_{\mu}$. In the context of Sec.~\ref{IV A}, $\tilde{\bm{A}}_{m} = \bm{Q}/(2e),~\tilde{A}_{m0}=0$, and $\delta\tilde{A}_{\mu}=(2e)^{-1}\partial_{\mu}\theta_v+A'_{\mu}$. A consequence of introducing the action $S_{\mathrm{J}}$ is that its fluctuation part cancels with the action stemmed from the $\mathcal{R}_1$ term in Eq.~(\ref{eq17}) or the $S_{\mathrm{I}}$ term in Eq.~(\ref{eq31a}). Otherwise, the extremum condition for Eq.~(\ref{GF}) would not be satisfied.

In this manner, now we have a principle for non-dissipative SC systems under supercurrent. However, the system we consider in this paper is a purely 2D superconductor, which is dissipative under a finite electric current due to the dissociation of vortex-antivortex pairs, causing a characteristic nonlinear resistivity. To deal with such a state, we refer to the Halperin-Nelson (HN) theory~\cite{Halperin}, where the electric response is described by using only the information of the SC state at the thermal equilibrium with zero current. In the same way, we can study the electric response based on the thermal equilibrium state under a supercurrent defined by Eq.~(\ref{GF}). The electric response of the dissipative state with a given supercurrent is addressed in this paper by the same logic as the HN theory~\cite{Halperin}. This is \textit{the main idea} of our work, allowing the formulation of nonreciprocal V-I characteristics. Following this principle, we extend the original HN theory to address the situation under the supercurrent where the total superfluid stiffness in Eq.~(\ref{eq28}) is anisotropic. The formulation of the nonlinear resistivity is developed in the next section.  

To simplify the analysis, we consider the vortex-free helical SC states, which are indeed the mean-field states of our model described in Sec.~\ref{II}. The free energy is given by Eq.~(\ref{eq10}). Since the parity symmetry is broken along the $x_2$ direction, a nonreciprocity accompanied by current flow is expected to occur in this direction. Hereafter, we focus on the external currents flowing in the $x_2$-direction. In this case, $\bm{j}_s$ satisfies 
\[       I_{s}\bm{e}_2 = L \int dx_1\, \bm{j}_s(\bm{x}). \tag{49} \label{Ij}            \]
Here, the integration is carried out over the length $L$ of the film. Note that for the thermodynamic argument, we have introduced $I_{s}$ which does not have the unit of electric current. The total electric current is given by $I_{s}$ multiplied by $2e/L$. As discussed in the last paragraph of Sec.~\ref{IV A}, we assume that the size of systems is much smaller than the Pearl's length $l_p$. In this limit, the supercurrent density is uniform across the cross-section of the system \cite{Clem}, and thus, $j_{s1}=0$ and $I_s = j_{s2} A_{\mathrm{s}}$. Since we consider the mean-field free energy in the presence of a uniform current along the $x_2$-direction, we have $F_{\mathrm{mf}}[\tilde{\bm{A}}_m] = F_{\mathrm{mf}}(Q_2\bm{e}_2)\equiv F_{\mathrm{mf}}(Q_2)$.  Therefore, the corresponding function to be minimized is
\[      G(Q_2) =  F_{\textrm{mf}}(Q_2)  - Q_2I_{s}, \tag{50}  \label{GFF}  \]
in which $F_{\mathrm{mf}}(Q_2)$ and $Q_2$ are related to the supercurrent $I_{s}$ as follows: 
\[     I_{s} = \pd{F_{\textrm{mf}}(Q_2)}{Q_2}.      \tag{51}  \label{Iy}        \]
From the discussion in Sec.~\ref{5V A}, we can replace $ f_{\mathrm{mf}}(\bm{Q})$ with the condensation energy $f_{\rm S}(\bm{Q})$.

\section{V-I characteristics\label{VII}}

In this section, we review the theory of nonlinear V-I characteristics in 2D superconductors and extend it for the purpose of the present study. 

\subsection{Generating Rate of Free Vortices\label{VII A}}

The theory of the resistive transition in 2D superconductors~\cite{Ambegaokar, Halperin} starts by considering the forces on a single vortex. There are an effective EM force $\bm{f}_E$, a Magnus force $\bm{f}_M$, a drag force $\bm{f}_D$, and a random force $\bm{f}_{\eta}$. The first three forces are given below:
\begin{align}
\bm{f}_{E} &= -\nabla_{\bm{r}^{(i)}} U \big( \bm{r}^{(1)},\dots,\bm{r}^{(N)} \big),  \tag{52a}  \label{Lor} \\
\bm{f}_{M} &= 2\pi  n_v \,\big( \bm{e}_3 \!\times \! \rho_s\bm{v}^{(i)}_{L}),   \tag{52b} \label{Mau} \\
\bm{f}_{D} &= -\mu^{-1}_v \, \bm{v}^{(i)}_{L}, \tag{52c} \label{Drag}         \end{align}
where $n_v=\pm1$, $\rho_s=2mJ_{B}$, $\bm{v}_{L}^{(i)}\equiv d\bm{r}^{(i)}/dt$ is the velocity of a vortex $i$, and $\mu_v$ is the vortex mobility. Here, $N$ is the number of vortices, and $U\big(\bm{r}^{(1)},\dots,\bm{r}^{(N)}\big)$ is the total vortex energy. Note that the Lorentz force was extracted from Eq.~(\ref{Mau}) to be included in Eq.~(\ref{Lor}) for later convenience. The expression of Eq.~(\ref{Mau}) can be tracked from Ref.~\cite{Sonin} easier than from Ref.~\cite{Ambegaokar}, and the expression of the drag force is brought from Ref.~\cite{Halperin}. Let us assume that the dynamics are overdamped, in other words, dissipative. By equating the summation of the forces to zero, we obtain the overdamped Langevin equation for a charged vortex as below:
\[    \D{\bm{r}^{(i)}}{t} = -\mu_v \nabla_{\bm{r}^{(i)}} U\big(\bm{r}^{(1)},\dots,\bm{r}^{(N)}\big) + \bm{f}^i_{\eta}(t). \tag{53} \label{Lagi}   \]
Here, $[ f^i_{\eta a}(t)f^j_{\eta b}(t') ]_{\eta} =2\eta T \delta_{ij}\delta_{ab}\delta(t-t')$, $[\cdots]_{\eta}$ denotes the average with respect to the fluctuation force, and the Magnus force in Eq.~(\ref{Mau}) was neglected because it is much smaller than the drag force. Now, we approximate $\nabla_{\bm{r}^{(i)} }U\big(\bm{r}^{(1)},\dots,\bm{r}^{(N)}\big)\simeq \nabla_{\bm{r}^{(i)}}\mathcal{U} (\bm{r}^{(i)}-\bm{r}^{(j)} )$, where $\mathcal{U} \big(\bm{r}^{(i)}-\bm{r}^{(j)} )$ is the total energy of a vortex $i$ and its antivortex $j$. We denote $\bm{r}=\bm{r}^{(i)}-\bm{r}^{(j)}$. Subtracting Eq.~(\ref{Lagi}) by the Langevin equation for antivortex $j$, we obtain the following equation:
\[         \bm{v}_L \equiv \D{\bm{r}}{t} = -2\mu_v \nabla_{\bm{r}} \,\mathcal{U} (\bm{r}) + \bm{f}_{\eta}(\bm{r},t), \tag{54} \label{Lag}     \]
where $[ f_{\eta a}(t)f_{\eta b}(t') ]_{\eta} =4\eta T \delta_{ab}\delta(t-t')$.  The total energy $\mathcal{U}$ of the vortex-antivortex pair with separation $|\bm{r}|$ can be given in the present case as follows:
\[      \mathcal{U} (\bm{r}) =  2\pi J_{B} \int_{\tilde{a}}^{\tilde{r}} \frac{dr'}{r'\eps(r')} -   \bm{f}_L\cdot\bm{r} -2\mu_0.  \tag{55} \label{U}  \] 
In the above equation, $\bm{f}_L$ is the Lorentz force given by $\bm{f}_L=2\pi n_v \, (\bm{j}_s\!\times\! \bm{e}_3)$ \cite{Sonin}, where $\bm{j}_s$ is a supercurrent density and $n_v$ is taken to be 1. Since we consider the total current flowing along the $x_2$-axis, $\bm{f}_L\cdot\bm{r} = 2\pi j_{s2}x_1$. 

The form of $\mathcal{U}(\bm{r})$ tells us that beyond a critical separation, namely $r_c$, which is obtained by the saddle point of Eq.~(\ref{U}), the force between the vortex and antivortex becomes repulsive, leading to the dissociation of them. The BKT theory is approximately valid at the separation less than $r_c$. In other words, the quasi-long-range interaction acts between vortices only within the length scale of $r_c$.

Following a standard analysis, we find a Fokker-Planck equation compatible with Eq.~(\ref{Lagi}) as follows:
\[   \pd{\Gamma (\bm{r},t)}{t} = -\nabla\lra{ -2\mu_v T e^{-\frac{\mathcal{U}(\bm{r})}{T} }\, \nabla \,   \Gamma(\bm{r},t) \, e^{\frac{\mathcal{U}(\bm{r})}{T} }    },  \tag{56} \label{FP}  \]
where $\Gamma(\bm{r},t)$ is the number density of vortex pairs dissociating in the separation $\bm{r}$. Ideally, we should solve Eq.~(\ref{FP}) to calculate the number density of free vortices $n_f$, which will be introduced later. However, the rigorous analysis is difficult because of the potential $\mathcal{U}$. In the original theory for the dissociation of vortex-antivortex pairs in superfluid films \cite{Ambegaokar}, $n_f$ was evaluated by using the stationary approximation method. In this scheme, first, we expand $\mathcal{U}$ around its stationary solution. From Eq.~(\ref{U}), the saddle point of $\mathcal{U}$ is $(x_{c1},x_{c2})$, where $x_{c2}=0$ and
\[   x_{c1} \equiv r_c = J(l_c)/j_{s2} = A_{\mathrm{s}}J(l_c)/I_{s}, \tag{57}  \label{xc}         \]
where $l_c = \ln \, (\tilde{r}_{c}/\tilde{a})$ is the critical scale factor, $\tilde{r}_{c} = r_{c}/\tilde{J}^{1/2}_{11}$, and $ j_{s2} = I_s/ A_{\mathrm{s}}$ was used. In the present method, this $r_{c}$ is the minimum distance at which a vortex-antivortex pair dissociates into two free vortices. Using this equation, $\mathcal{U}$ can be approximated as 
\[   \mathcal{U}(\bm{r})\simeq  \mathcal{U}_0 - \frac{\pi J(l_c)}{r^2_{c}} \lra{ (\delta x_1)^2  - \frac{\tilde{J}_{11}}{\tilde{J}_{22}} (\delta x_2)^2 },       \tag{58} \label{sae}  \]
where $\mathcal{U}_0 = \mathcal{U}(x_{c1},0)$, $\delta x_1 = x_1-x_{c1}$, and $\delta x_2 = x_2-x_{c2}$. In obtaining the above equation, the derivative of $J(l)$ with respect to $\delta x_1$ and $\delta x_2$ is neglected as in Ref.~\cite{Ambegaokar}. For later convenience, we introduce $\mathcal{U}_{1} =  -\pi J(l_c)(\delta x_1)^2/r^2_{c}$ and $\mathcal{U}_2 = \pi J(l_c)\tilde{J}_{11}(\delta x_2)^2/(r^2_{c}\tilde{J}_{22})$, and then $\mathcal{U}(\bm{r})\simeq \mathcal{U}_0 + \mathcal{U}_{1} + \mathcal{U}_{2}$.

Next, let us focus on the term in the square bracket of Eq.~(\ref{FP}). This term represents a diffusive current density $\bm{j}_d(\bm{r})$. With this and the normalization of $\Gamma(\bm{r},t)$, the number of separating vortex pairs per unit film area per unit time is given by the quantity $R_v \equiv \int d x_2\,j_{d1}(r_c,x_2)$. The integral is taken on the $x_2$-axis instead of the $x_1$-axis because the pure Lorentz force makes the vortices diffuse in the direction perpendicular to the supercurrent. Since we are interested in a steady state in which $\Gamma (\bm{r},t)$ is not dependent on time, we obtain $\nabla \cdot \bm{j}_{d}=0$ from Eq.~(\ref{FP}). Let us now make the ansatz that $j_{d2}=0$. Then, it turns out that $j_{d1}$ must not depend on $\delta x_1$, and $\Gamma(\bm{r})$ in the expression of $j_{d2}$ has to cancel out all terms depending on $x_1$ from $\exp\,(\mathcal{U}(\bm{r})/T)$. From the definition of $R_{v}$ and the expression of $j_{d1}$, we obtain the following:
\begin{align*}        
R_v \, e^{\mathcal{U}_1/T} &= -2\mu_v T e^{\mathcal{U}_1/T} \int d\delta x_2 \,  e^{-\mathcal{U}/T} \partial_1 \big[ \Gamma(\bm{r}) \, e^{\mathcal{U}/T} \big] \\
&= -2\mu_v T  e^{-\mathcal{U}_0/T} \bigg(\frac{\tilde{J}_{22} r^2_c}{\tilde{J}_{11} K(l_c)} \bigg)^{1/2} \partial_1 \big[ \Gamma(\bm{r}) \, e^{\mathcal{U}/T} \big].  \tag{59}
\end{align*}
In the derivation of the last equation, we used the fact that there is no dependence on $\delta x_2$ in the square bracket. By further integrating both sides of the above equation with respect to $\delta x_1$ from vortex diameter $a$ to $\infty$, we obtain the following expression:
\[       R_v \simeq 2\mu_v T e^{-\mathcal{U}_0/T} \bigg(\frac{\tilde{J}_{22} }{\tilde{J}_{11} } \bigg)^{1/2} \big[ \Gamma(\bm{r}) \, e^{\mathcal{U}/T} \big]_{\delta x_1=a}.     \tag{60}  \label{Rme}           \]
Here, the Gaussian integration was approximated by extending the lower limit of the integration to $-\infty$, and we used the fact that $ [ \Gamma(\bm{r}) \exp\, \big({\mathcal{U}/T} ) ]_{\delta x_1\rightarrow\infty}=0$. In Ref.~\cite{Ambegaokar}, the term $ [ \Gamma(\bm{r}) \, \exp\, (\mathcal{U}/T) ]_{\delta x_1=a}$ is given by $a^{-4}$ without an explicit explanation. To show this, we follow the original theory of the escape rate by Chandrasekhar~\cite{Chandrasekhar}. At separations less than $r_c$, vortices tend to form pairs rather than dissociate into isolated vortices and are approximately in thermal equilibrium. Thus, the number density $\Gamma(\bm{r})$ for those pairs is approximately given as $\Gamma(\bm{r}) = C \exp\, ( -\mathcal{U}/T )$, in which the Lorentz force term $-\bm{f}_L\cdot\bm{r} = - 2\pi j_{s2}x_1$ is removed from $\mathcal{U}$ in Eq.~\eqref{U}. To find the normalization factor $C$, we consider a pair with internal separation $a$. Following Ref.~\cite{Ambegaokar}, we have 
\[       Ca^2\int_0^{a_1} dx_1 \int_0^{a_2} dx_2    \, e^{ -\mathcal{U}/T} = 1.       \tag{61}           \]
Since the first and second terms of the r.h.s. in Eq.~(\ref{U}) are removed and neglected, respectively, in the distance less than $a$, $\mathcal{U}$ is given by $-2\mu_0$. From the scale transformation discussed in Sec.~\ref{IV B}, we obtain the normalization factor as $C \simeq a^{-4}$. Consequently, we have $ [ \Gamma(\bm{r}) \, \exp\, (\mathcal{U}/T) ]_{\delta x_1=a}\simeq a^{-4}$. Substituting this into Eq.~(\ref{Rme}), we reach the following expression:
\[       R_v \simeq \frac{2\mu_v T}{a^4}  \, e^{-\mathcal{U}_0/T} \bigg(\frac{\tilde{J}_{22} }{\tilde{J}_{11} } \bigg)^{1/2} .     \tag{62}  \label{Rxy}           \]

In the steady state, the generation rate of free vortices $R_v$ equals the recombination rate of free vortices to a vortex-antivortex pair $R_e$. The latter is given by $R_e = [v_{L1}]_{\eta} \sigma_c n^2_f$, where $\sigma_c$ is the cross-section of the vortex-antivortex pair combination and $n_f$ is the number density of free vortices. For simplicity, we approximate that the distance between isolated vortices is sufficiently large so that the inter-vortex interaction can be neglected. In this manner, by averaging Eq.~(\ref{Lagi}) with $\eta$, we have $[v_{L1}]_{\eta} \simeq 2\pi\mu_v j_{s2}$. For $\sigma_c$, it is clear that its value is of order $r_c$, and thus we adopt $\sigma_c\simeq r_c$. Combining these expressions with Eq.~(\ref{xc}), we obtain $R_e \simeq 2\pi \mu_v J(l_c) \, n^2_f$. From the condition $R_v=R_e$, the number density of free vortices $n_f$ is evaluated as below:
 \[         n_f =   \frac{ \lr{\tilde{J}_{22}/\tilde{J}_{11}}^{1/4} }{\lra{\pi a^4 K(l_c)}^{1/2}} \exp\lra{-\mathcal{U}_0/(2T)}.  \tag{63} \label{nfr}\]
From Eq.~(\ref{yl}), the exponential of the above equation can be rewritten as $Y(l_c)  \lr{\tilde{a}/\tilde{r}_c}^{2}  \exp\lra{\pi K(l_c)}$. Combining this with the above equation, we obtain the following expression:
 \[         n_f =   \frac{  Y(l_c) \exp\lra{\pi K(l_c)}}{r^2_c \lra{\pi  K(l_c)}^{1/2}} \lr{\frac{\tilde{J}_{11}}{\tilde{J}_{22}} }^{1/4}.  \tag{64} \label{nf}\]

It should be noticed that not only the dependence of various parameters on the supercurrent or Cooper pair momentum but also the factor due to the anisotropy $\tilde{J}_{11}/\tilde{J}_{22}$ do not appear in the original theory for vortex-antivortex dissociation~\cite{Ambegaokar, Halperin}. This factor comes from the prescription in Sec.~\ref{IV B}.

\subsection{Nonlinear Resistance\label{VII B}}

Using the number density of free vortices evaluated in the preceding section, we derive the expression of DC resistance associated with the generation of free vortices. Instead of using the Josephson relation as in the original HN theory~\cite{Halperin}, here we show alternative derivation to obtain the following Eq.~(\ref{Rsi}) for the resistance. 
We begin with the well-known fact that the motion of a vortex along the direction perpendicular to the supercurrent causes the electric field $\bm{E}' = - \bm{v}_L \!\times\! \bm{B}$, where $\bm{B}$ is a magnetic field in the vortex core, parallel to the supercurrent exerting on the vortex~\cite{Kim}. Approximating that the magnetic field is uniform over the vortex core, we have $e\bm{E}'^{(i)}=\pi n_v\bm{e}_3\!\times\!\bm{v}^{(i)}_{L}/A_{\mathrm{s}}$ for a free vortex $i$. Indeed, this is how the Magnus force in Eq.~(\ref{Mau}) is derived. The contribution to voltage from the force in the $x_1$-direction was neglected when considering the motion of a vortex in a perpendicular direction. But now, we consider the contribution from the force in the $x_2$-direction. Since the number of free vortices is $n_fA_{\mathrm{s}}$, the total average voltage $\Delta V=[\sum_i E'^{(i)}_{2} L ]_{\eta}$ across the two ends of the film in the $x_2$-direction is   
\[    \Delta V/L = \pi \, n_f v_{L1}/e.     \tag{65} \label{V}    \]
Inserting $[v_{L1}]_{\eta} \simeq 2\pi\mu_v j_{s2}$, which was derived in the previous section, into the above equation, the DC resistance $R_s$ can be expressed as follows:
\[   R_s =  \Delta V/\mathcal{I}_s = \pi^2 \mu_v n_f /e^2,   \tag{66} \label{Rsi}           \]
where $\mathcal{I}_s=2ej_{s2} L$ is the electric supercurrent. 

The unknown quantity is now only the vortex mobility $\mu_v$. From the Bardeen-Stephen theory~\cite{Bardeen}, it is given as
\[\mu_v=2e^2\xi^2_{\textrm{GL}}/(\pi\sigma_N), \tag{67} \label{vom} \]
where $\sigma_N$ is the 2D normal conductivity and $\xi_{\textrm{GL}}$ is the GL coherence length. This formula is derived for dirty superconductors~\cite{Halperin}, in which the impurity scattering time of normal electrons in the vortex cores $\tau_e$ satisfies the condition $\Delta_0 \tau_e\ll 1$ with $\Delta_0$ being the amplitude of gap function at $T=h=0$. The condition can also be rewritten as $\xi_e\ll \xi_0$, where $\xi_e= v_{\rm F}\tau_e$ is the mean free path of electrons and $\xi_0$ is the BCS coherence length at $T=h=0$. However, it has been shown that Eq.~(\ref{vom}) is approximately valid for moderately clean superconductors, where the condition $\Delta^2_0/E_{\rm F}\ll 1/\tau_e\ll \Delta_0$ is satisfied, in studies of vortex flow under a perpendicular magnetic field~\cite{Kopnin, Skvortsov}. Although we do not include any impurity effect in the model, qualities specifying the SC state are expected not to change significantly if the condition $\Delta \tau_e \gg 1$ holds. Therefore, we adopt Eq.~(\ref{vom}) with replacing $\xi_{\textrm{GL}}$ by the BCS coherence length $\xi$.

From the above discussion and Eqs.~(\ref{nf}), (\ref{Rsi}), and \eqref{vom}, we then obtain the DC resistance $R_s$ as follows:
\begin{align*}   
R_s/R_N &= 2\pi \xi^2 n_f \\
&= \frac{2\pi\xi^2 }{r^2_c} \frac{  Y(l_c) \exp\lra{\pi K(l_c)}}{ \lra{\pi  K(l_c)}^{1/2}} \lr{\frac{\tilde{J}_{11}}{\tilde{J}_{22}} }^{1/4}, \tag{68} \label{Rsf}           
\end{align*}
where $R_N = \sigma^{-1}_N$ is the normal resistance. The original HN theory considered only Region I where we obtained Eqs.~(\ref{XX1}) and (\ref{yl1}). However, in the vicinity of the BKT transition, the bare parameters can enter Region II, where $X(l)$ and $Y(l)$ are given by Eqs.~(\ref{XXX2}) and (\ref{yl2}). In that case, we will use the expression of $K(l_c)$ in Region II to compute the resistance.

As seen in Eq.~(\ref{Rsf}), there is no restriction that the resistance due to free vortices is smaller than the normal resistance. To the best of our knowledge, theory based on the vortex-antivortex dissociation picture does not satisfactorily formulate the crossover to the normal state as the current increases. To satisfy this, we may regard the conductivity $\sigma_s\propto R^{-1}_s$ as the contribution from the supercurrent affected by the vortex-antivortex dissociation, and the normal conductivity $\sigma_N$ should be added to it~\cite{Mondal, Benfatto}. If this is true, the total resistance $R_{\mathrm{tot}}$ becomes 
\[  R^{-1}_{\mathrm{tot}} = R^{-1}_s + R^{-1}_N.        \tag{69} \label{interpolation}         \]
We can interpret this equation by considering that not only the supercurrent $\mathcal{I}_s$ but also the normal current $\mathcal{I}_N$ flows. 
Since we have not taken into account the normal current in the mean-field analysis, our approach is valid when the condition $\mathcal{I}_N/\mathcal{I}_s=R_s/R_N \ll 1$ holds. The limit of $R_s/R_N$ below which our approach is valid has not been clarified. However, as will be seen in the next section, intriguing nonlinear transport phenomena are observed even for $R_s/R_N\sim10^{-2}$. It is expected that our formulation is justified for such a small $R_s/R_N$. In the following sections, we also show the results in the regime where the nonlinear resistance $R_s$ becomes comparable to $R_N$. Although the correctness of our results in this regime is not guaranteed, it will be shown that our results reproduce some aspects of the experiments. For brevity, we calculate the total resistance $R_{\mathrm{tot}}$ approximately by the following equation:
\[R_{\mathrm{tot}} \simeq \left\{ \begin{array}{lll}
R_s & \text{when} & R_s\leq R_N\\
R_N & \text{when} & R_s>R_N. \end{array} \right. \tag{70} \label{approx} \]
An advantage of this simplification is that the crossover to the normal state is clearly defined. See more details at the beginning of Sec.~\ref{VIII}.
 
Noted that the scenario of nonlinear resistivity due to the vortex-antivortex pair dissociation is valid only at finite temperatures. Recently, it has been argued that the resistance of a finite width sample near zero temperature is dominated by the quantum phase slip \cite{Konig}. Since we are interested in the phenomena at finite temperatures, it is justified to neglect this mechanism.

Let us comment on the formula for the nonlinear resistivity in the original HN theory~\cite{Halperin}. The formula is given by neglecting the Cooper pair momentum dependence of parameters in Eq.~(\ref{Rsf}) and using only the information of the zero current states. Moreover, the vortex core radius is set to the coherence length of the zero current state at the BKT transition temperature. Therefore, nonreciprocal V-I characteristics analogous to the SD effect cannot be expected from their formula, even when the time-reversal symmetry and inversion symmetry are broken. Below, we refer to the formula obtained by those approximations as the original HN formula. Our formulation leading to Eq.~\eqref{Rsf} extends the HN theory by removing the above assumptions and enables us to calculate the nonreciprocal V-I characteristics.

In the original HN theory \cite{Halperin}, the analytic formulae were derived in the limits $l_c\sqrt{|C_0|}\ll 1$ and $l_c\sqrt{|C_0|}\gg 1$, resulting in limitation of the applicability of the formulae. We refer to them from now on as asymptotic expressions. Although we numerically calculate the nonlinear resistance by using the extended formula in Sec.~\ref{VIII A}, the asymptotic expressions in the original HN formula give important relations. When the supercurrent is applied at the BKT transition temperature, $l_c$ is finite while $C_0=0$. Thus, the asymptotic expression in the limit $l_c\sqrt{|C_0|} \ll1$ is valid. In this limit, the resistance obtained by the HN theory, $R_{\mathrm{HN}}$, is 
\[  R_{\mathrm{HN}}\propto x_I(I_s/I_0)^{2+x_I/2},     \tag{71} \label{HN}   \] 
where $x_I=1/\ln\, (I_0/I_s)$, $I_0=A_{\mathrm{s}}J_0(l_c)/\xi_c$, and $J_0(l_c)$ denotes the renormalized superfluid stiffness in the zero current state at the scale $l_c$. At low currents, $l_c\rightarrow\infty$, and we can show that $R_{\rm HN}\propto I^2_s$. This is known as a universal relation. Our formula, Eq.~(\ref{Rsf}),  naturally satisfies this relation. At high currents, the HN theory further approximates as $J_0(l_c) \simeq J_0(l_c=0)$, making $I_0$ independent of $I_s$. Consequently, the resistance $R_{\mathrm{HN}}$ is convex to the current in the log scale as long as $I_s<I_0$. Without the above approximations, it is not easy to predict the behavior of $R_{\mathrm{HN}}$ analytically. However, as shown in Sec.~\ref{VIII A}, the V-I characteristics numerically calculated from the original HN formula at the zero field BKT transition temperature shows a convex curve in the log scale in the range $R_{\mathrm{HN}}<R_N$, consistent with the analytic expressions.

In the other limit $l_c\sqrt{|C_0|}\gg 1$, which realizes at temperatures below the BKT transition when the current is sufficiently low, the resistance in the HN theory reads as
\[     R_{\mathrm{HN}}\propto  x_T (I_s/I_0)^{2+x_T/2}, \tag{72} \label{HN1}            \]
where $x_T= 2\pi K_{R}-4$, and $K_{R}=J_{R}/T$ has been introduced in Sec.~\ref{IV C} with the full renormalized superfluid stiffness $J_{R}$. We again notice that the HN theory adopts the value of zero current states. It can be shown that the extended formula Eq.~(\ref{Rsf}) satisfies the power law $R_s\propto I_s^{2+x_T/2}$ in the same limit.

\section{Numerical Results of V-I Characteristics\label{VIII}}

In this section, we show the nonlinear V-I characteristics near the BKT transition line. The nonlinear resistance is calculated by using Eqs.~(\ref{Rsf}) and (\ref{approx}) with the parameters from Sec.~\ref{5V A}. In the following Secs.~\ref{VIII A} and \ref{VIII B}, we show the results at zero magnetic field, where the V-I characteristics are reciprocal. In Secs.~\ref{VIII C} and \ref{VIII D}, we present the nonreciprocal V-I characteristics at finite magnetic fields. We discuss the finite size effect in Sec.~\ref{VIII D}.

The crossover from the BKT critical region to the normal region occurs when the V-I characteristics become linear. Based on the assumption Eq.~(\ref{approx}), the crossover can be defined by the condition $R_s=R_N$. We refer to the voltage and current at the crossover as the crossover voltage and the crossover current, respectively. Furthermore, a current flowing in the positive (negative) direction is referred to as a forward (backward) current. The forward (backward) voltage and resistance are defined in the same way.


\subsection{Reciprocal V-I Characteristics at Zero Magnetic Field\label{VIII A}}

Before showing the calculation results, let us mention some experimental facts. In the early V-I measurements in purely 2D superconductors \cite{Hebard, Epstein, Fiory, Garland}, the main focus was on the sufficiently low current regime. In this regime, as discussed at the end of the previous Sec.~\ref{VII B}, the voltage $V$ and supercurrent $I_s$ near the zero field BKT transition temperature $T_{\rm c}$ obey the HN scaling law $V\propto I^{3+\alpha}_s$, where $\alpha>0$ at $T<T_{\rm c}$ and $\alpha=0$ at $T=T_{\rm c}$. In recent years, the measurements in the high current regime have also been performed \cite{Lin, Venditti, Saito, Hua, Liu, Weitzel}. The results show that the V-I characteristics deviate from the scaling law in the high current regime. Roughly speaking, there are two experimental observations beyond the HN scaling law in the high current regime at temperatures slightly lower than $T_{\rm c}$. At temperatures close to $T_{\rm c}$, the V-I characteristics show a convex behavior and change to be concave as increasing the current toward the crossover to the normal phase. At sufficiently low temperatures, the concave part shrinks, leaving only the convex part. These behaviors of deviation from the HN scaling law may be explained by the heating effect on the sample~\cite{Weitzel}. Moreover, in the intrinsically inhomogeneous samples, another mechanism can produce V-I characteristics similar to those of the vortex-antivortex pair dissociation mechanism \cite{Venditti}. In the following, we focus on homogeneous superconductors and show that the convex and concave behaviors observed in experiments can also be described in a unified way within the extended framework of the HN theory, namely, Eq.~\eqref{Rsf}.

\begin{figure}[t]
\centering

\includegraphics[width=8.5cm]{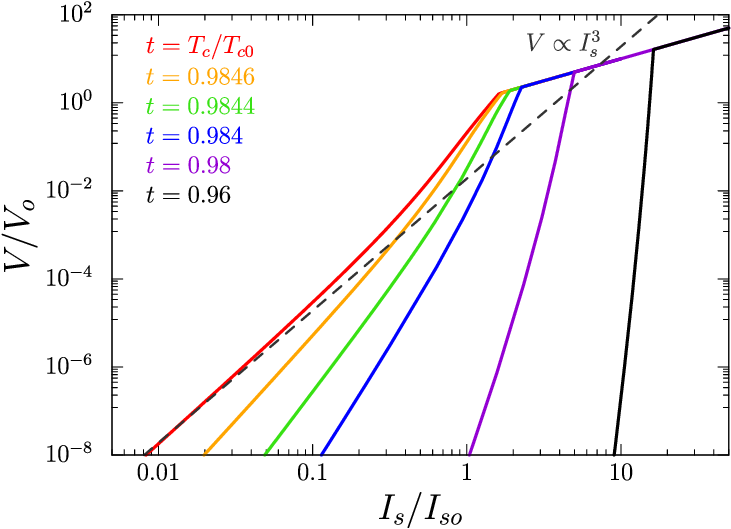}

\caption{The nonlinear V-I characteristics at zero magnetic field for various temperatures below the zero-field BKT transition temperature $T_{\rm c}$. The V-I characteristics are reciprocal in this case. The red, orange, green, blue, violet, and black lines correspond to reduced temperatures $t=0.98467(5) (=T_{\rm c}/T_{\rm c0}),~0.9846,~0.9844,~0.984,~0.98$, and 0.96, respectively. The dashed grey line shows the universal relation $V\propto I^3_s$. The current and voltage are scaled by $I_{so} = N(0)T^2_{\rm c0}/p_{\rm F}$ and $V_o = I_{so}R_N$, respectively.} 

\label{fig8}
\end{figure}

In Fig.~\ref{fig8}, we show the numerical results of V-I characteristics in the absence of the magnetic field.  As the temperature decreases, the curves shift to the large current side, resulting in small resistance and large crossover voltage. At the same time, the curves become sharper, and their concave part near the crossover to the normal state becomes narrower. Thus, our calculations reveal the V-I curves whose convexity is consistent with the experimental results described at the beginning of this section. An exceptional result has been reported in Ref.~\cite{Liu}, in which the concave part is absent even at temperature near the BKT transition temperature.

The presence of the convex and concave parts in the V-I curves reveals the deviation from the HN scaling law. The deviation appears at a current much smaller than the crossover current. For example, at the zero-field BKT transition temperature (red line in Fig.~\ref{fig8}), the deviation from the scaling law $V\propto I^3_s$ can be observed even at $I_s\sim 0.1 I_{so}$, which corresponds to the resistance $R_s /R_N\sim 3\times 10^{-4}$. At finite magnetic fields, as we discuss in Sec.~\ref{VIII C}, the V-I curve can also become concave to the current, depending on the current direction, even when the current is much smaller than the crossover current.

When the supercurrent flows through an ideal superconductor, the superconducting part of the circuit can be approximated as being in thermal equilibrium. Thus, a mean-field description of the current-carrying state, such as the helical state, is allowed in superconductors. Even when the zero current state is not the helical state but the stripe state, the mean-field superconducting state approaches to the helical state as the current increases~\cite{Aoyama}. The 2D superconductors are affected by dissipation when any finite electric current flows, making it difficult to accurately describe the state. However, if the dissipation rate or resistance is sufficiently small, it is natural to expect the system to be described as a nearly equilibrium current-carrying state suffering from vortex-antivortex dissociation. Therefore, our formulation of purely 2D superconductors would require $R_s/R_N\ll1$, the same condition as that discussed around Eq.~\eqref{interpolation} in the previous section. Thus, as the current increases in 2D helical superconductors while maintaining the condition $R_s/R_N\ll1$, it can be interpreted that the Cooper pair momentum increases, as in the mean-field theory. As we showed above, deviation from the HN scaling law is observed in this region.

\begin{figure}[t]

\begin{minipage}{0.49\hsize}
\subfloat[\label{wj-8}]{\includegraphics[width=4.03cm]{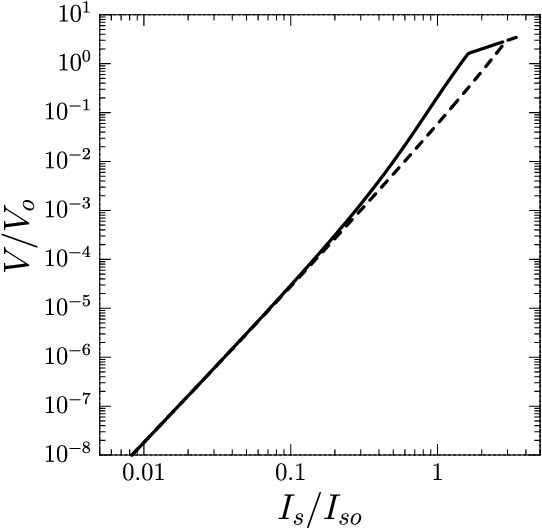}}
\end{minipage}
\begin{minipage}{0.49\hsize}
\subfloat[\label{wj-9}]{\includegraphics[width=4.2cm]{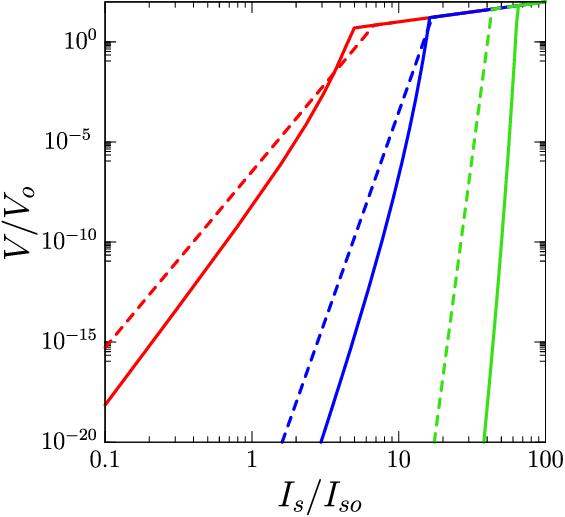}}
\end{minipage}

\captionsetup{font=normal,justification=raggedright,singlelinecheck=false}
\caption{Comparison of our extended formulation [Eq.~\eqref{Rsf}, solid lines] and the original HN theory [dashed lines]. The nonlinear V-I characteristics at zero magnetic field are compared (a) at the zero-field BKT transition temperature $T_{\mathrm{c}}$ while (b) below $T_{\mathrm{c}}$. The red, blue, and green lines in the panel (b) show the V-I curves at $t=0.98,~0.96$, and 0.9, respectively. }

\label{fig9}
\end{figure} 

Let us compare the results of the extended HN formula in Fig.~\ref{fig8} with the original HN formula [see Sec.~\ref{VII B}]. As shown in Fig.~\ref{fig9}, at the BKT transition temperature $T_{\rm c}$, the original HN formula gives a larger crossover current than our formula, and an almost straight line implies the scaling law in the whole current region below the crossover current. However, the V-I curve is slightly concave before crossover to the normal region, although it is difficult to see with the naked eye.  Thus, the original HN formula is consistent with the asymptotic behavior of Eq.~(\ref{HN}). On the other hand, the V-I curve obtained based on our approach shows a convex behavior in a wide region and becomes concave near the BKT critical-to-normal crossover current. In various experiments~\cite{Lin, Venditti, Saito, Hua, Liu, Weitzel}, the V-I curves near the BKT transition are concave to the current in the log scale in the high current region below the crossover current. Therefore, our approach, which extends the HN theory, can reproduce the V-I curves near the BKT transition temperature in better agreement with the experiments than the original HN formula.

Below the BKT transition temperature, for instance, at $t=0.98$, it is hard to see the convex part in the V-I curve calculated from the original HN formula, and the voltages in the low current region are significantly different from those obtained by our formula. The difference arises because the original HN theory assumes the vortex core radius at the zero field BKT transition temperature $T_{\rm c}$. In contrast to the original HN theory, our approach shows the convex V-I characteristics. In some experiments on the 2D superconducting films, e.g. Refs.~\cite{Lin, Venditti, Liu}, the V-I curves at temperatures lower than $T_{\rm c}$ have a clear convex part at not-too-high currents. Therefore, our approach can reproduce the experimentally observed behavior of the nonlinear resistivity better than the original HN theory.

\subsection{Vortex Core Size Dependence\label{VIII B}}

\begin{figure}[t]

\begin{minipage}{0.49\hsize}
\subfloat[\label{fig10a}]{\includegraphics[width=4.03cm]{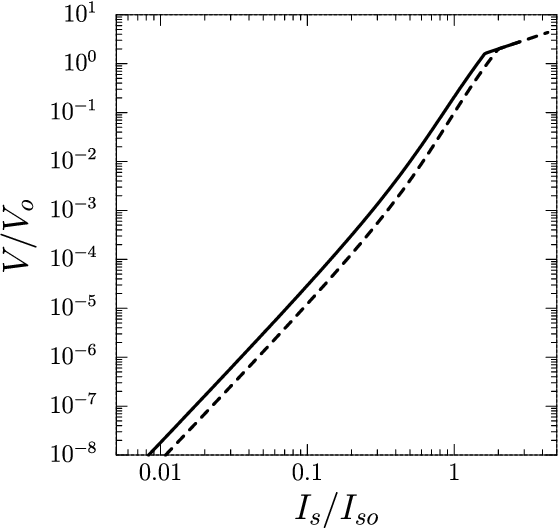}}
\end{minipage}
\begin{minipage}{0.49\hsize}
\subfloat[\label{fig10b}]{\includegraphics[width=4.2cm]{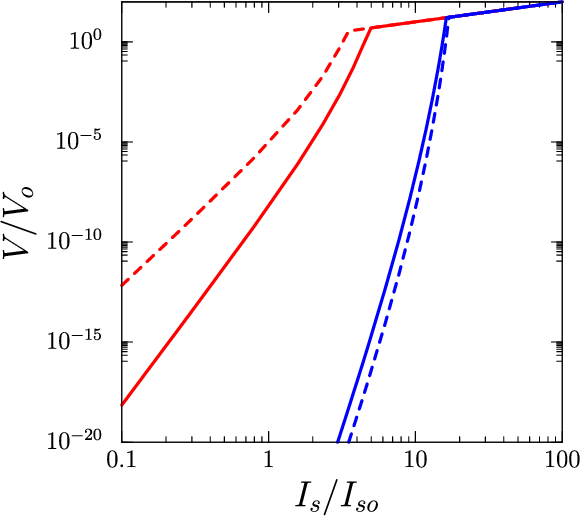}}
\end{minipage}

\captionsetup{font=normal,justification=raggedright,singlelinecheck=false}

\caption{The V-I characteristics at zero magnetic field calculated with the vortex core radius of the BCS and GL coherence lengths. The solid lines show the V-I curves calculated with the vortex core radius of the BCS coherence length, while the dashed lines show those with the GL coherence length. (a) The results at each zero-field BKT transition temperature when the BCS and GL coherence lengths are adopted. (b) The red and blue lines show the V-I curves at $t=0.98$ and 0.96, respectively.}

\label{fig10}
\end{figure}

Until now, we have assumed the BCS coherence length $\xi$ for a vortex core size. In this section, we assume the GL coherence length $\xi_{\mathrm{GL}}=0.74\xi$ \cite{Fetter} for a vortex core radius to examine how the results depend on the vortex core size. We find two effects. First, the zero-field BKT transition temperature slightly decreases by 0.34\% from the result shown in Sec.~\ref{IV}. This is to be expected because more vortices are excited as the radius of vortex cores becomes smaller. As shown in Fig.~\ref{fig10a}, the V-I curve remains unchanged qualitatively.

Second, as shown in Fig.~\ref{fig10b}, the crossover voltage and current turn out to be larger than the results using the BCS coherence length at low temperatures while it is smaller at temperatures close to the BKT transition temperatures. The origin of the decrease of the crossover voltage and current is unclear. However, the reason for the increase at lower temperatures can be explained. We find from the numerical data that the effect comes from the reduction of the term $\exp\, (-2l_c\sqrt{|C_0|})$ in Eq.~(\ref{yl1}). Further details can be found in Appendix \ref{AppendixB}.

From the above results, we conclude that only quantitative changes are caused by the choices of vortex core radius, namely the BCS and GL coherence lengths. 

\subsection{Nonreciprocal V-I Characteristics in Finite Magnetic Fields\label{VIII C}}

\begin{figure}[b]
\captionsetup{font=normal,justification=raggedright,singlelinecheck=false}
\begin{minipage}{0.49\hsize}
\subfloat[$t=0.7$\label{fig11a}]{\includegraphics[width=4.2cm]{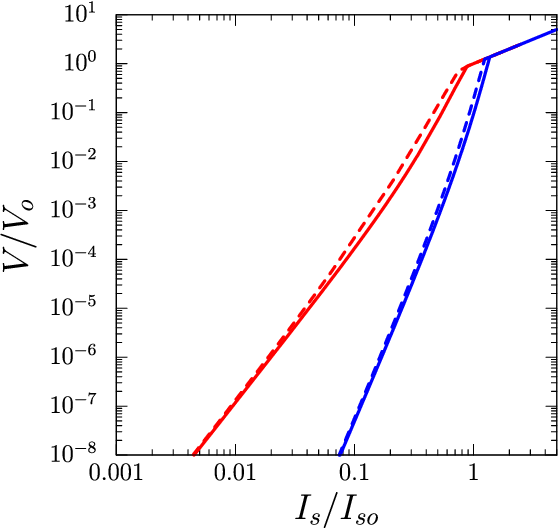}}
\end{minipage}
\begin{minipage}{0.49\hsize}
\subfloat[$t=0.6$\label{fig11b}]{\includegraphics[width=4.25cm]{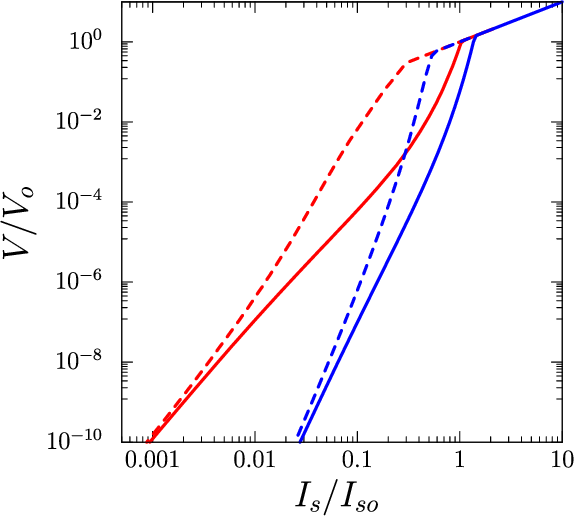}}
\end{minipage}
\begin{minipage}{0.49\hsize}
\subfloat[$t=0.5$\label{fig11c}]{\includegraphics[width=4.2cm]{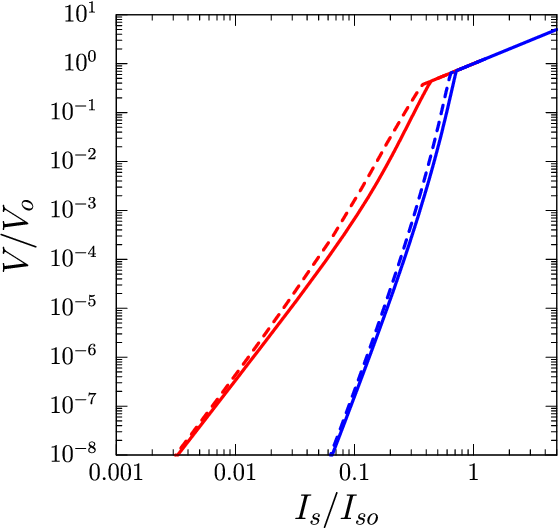}}
\end{minipage}
\captionsetup{font=normal,justification=raggedright,singlelinecheck=false}
\caption{The nonlinear V-I characteristics at finite magnetic fields for various temperatures. The red curves are obtained at the temperatures and magnetic fields on the BKT transition line, while the magnetic fields of the blue curves are slightly lower than those of the red ones by $0.002 \, T_{\mathrm{c0}}$. The solid and dashed curves show the V-I characteristics for the forward and backward currents, respectively.}

\label{fig11}
\end{figure}

In this section, we show the nonreciprocal V-I characteristics in the magnetic fields, which is analogous to the SD effect. Figure \ref{fig11} shows the results at and slightly below the BKT transition temperature. When we fix the magnitude of the electric current, the forward voltage is smaller than the backward voltage, revealing the nonreciprocal V-I characteristics. In particular, the V-I curves at $t=0.6$ exhibit strong nonreciprocity in the DC resistance and the crossover current and voltage. The strong nonreciprocity is obtained near the crossover line of helical SC states, discussed in Sec.~\ref{5V C}. This feature is analogous to the SD effect in the mean-field theory: the magnitude of the SD effect is drastically increased in the crossover region~\cite{Daido1,Ilic}. Reflecting this feature of bulk superconductors, purely 2D superconductors show strong nonreciprocity associated with crossover in the helical SC states.

It should be noted that amplitude-mode renormalization is essential for the nonreciprocal V-I characteristics near the crossover line. For instance, from the numerical data at $t=0.6$ in Fig.~\ref{fig11b}, we find that near the BKT critical-to-normal crossover under the forward current $\tilde{J}_{22}$ reaches the global maximum as a function of the Cooper pair momentum $Q_2$. As described below Eq.~(\ref{eq25}), $\tilde{J}_{\mu\nu}~(J_{\mu\nu})$ denotes the $\mu\nu$ component of the superfluid stiffness where the amplitude-mode fluctuation is (not) renormalized. The difference between $J_{22}$ and $\tilde{J}_{22}$ is significant when the latter reaches its global maximum as shown in Fig.~\ref{fig1}. Therefore, the amplitude-mode fluctuation plays a particularly important role in the nonlinear V-I characteristics around the crossover region of helical SC states.

\begin{figure}[t]

\captionsetup{font=normal,justification=raggedright,singlelinecheck=false}

\begin{minipage}{\hsize}
\subfloat[Forward direction\label{fig12a}]{\includegraphics[width=7.2cm]{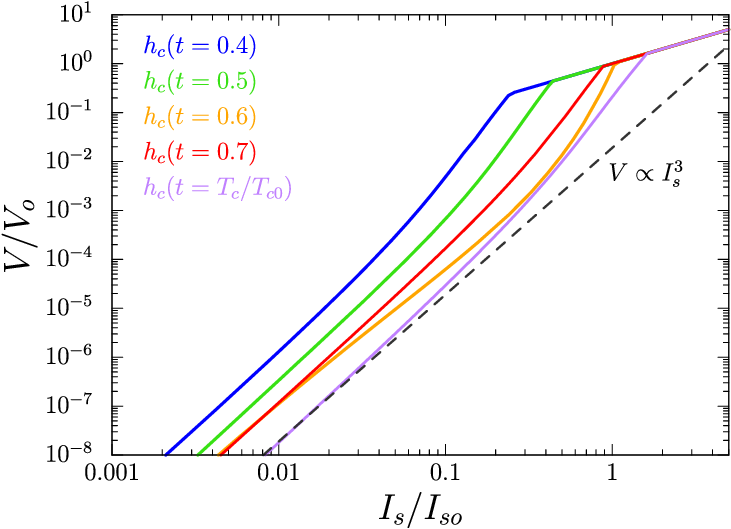}}
\end{minipage}
\begin{minipage}{\hsize}
\subfloat[Backward direction\label{fig12b}]{\includegraphics[width=7.2cm]{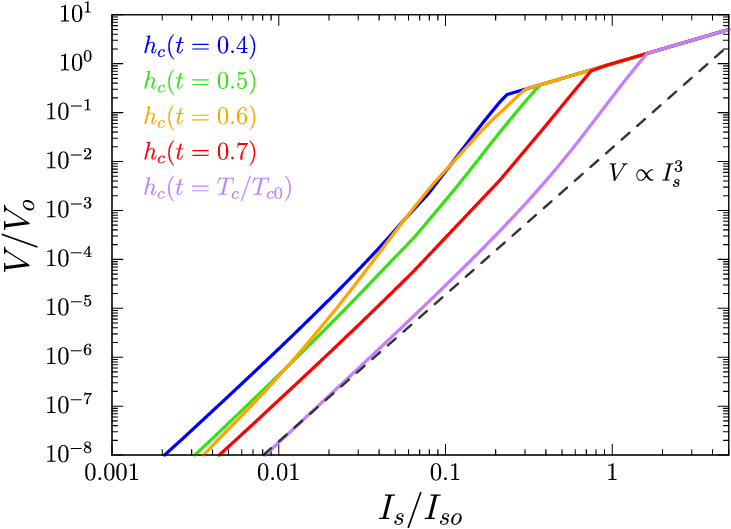}}
\end{minipage}

\captionsetup{font=normal,justification=raggedright,singlelinecheck=false}
\caption{The V-I characteristics in the (a) forward and (b) backward current directions at the BKT transition temperatures. The blue, green, orange, red, and purple lines show the V-I curves at reduced temperatures $t=0.4,~0.5,~0.6,~0.7$, and $T_{\mathrm{c}}/T_{\mathrm{c0}}$, respectively. The magnetic field increases with decreasing $t$ following the BKT transition line. The dashed lines show the universal scaling relation $V\propto I^3_s$. All the V-I curves, even for $t=0.6$, are parallel to the dashed line at sufficiently low currents, consistent with the universal relation.}

\label{fig12}
\end{figure}

Next, let us look at the shape of the V-I curves. As shown in Fig.~\ref{fig11}, the convexity of the V-I curves has a directional dependence on the currents. In particular, the V-I curves near the crossover of helical SC states, namely at $t=0.6$, show different convexity near the BKT critical-to-normal crossover. When the current flows in the forward direction, the convex curve appears in the wide range of current, while the concave part becomes broadened if the current flows in the opposite direction. As stated in Sec.~\ref{VIII A}, these features with significant nonreciprocity appear at sufficiently low currents, showing an intriguing phenomenon arising from the helical crossover.

Now, we rearrange the V-I curves in Fig.~\ref{fig11} and for other parameters on the same graph in Fig.~\ref{fig12}. We find the following two intriguing features of the V-I characteristics along the BKT transition line. First, the temperature (or magnetic field) dependence is non-monotonic and nonreciprocal. When the current flows in the forward direction, as shown in Fig.~\ref{fig12a}, the V-I curve shifts to the lower current region with increasing the magnetic field, except for the case of $t=0.6$ near the crossover of helical SC states. A similar behavior appears when the current flows in the opposite direction, but the non-monotonic behavior is observed around $t=0.5$ in Fig.~\ref{fig12b}. This is due to the enhanced nonreciprocity around the crossover line of helical SC states. The BKT critical-to-normal crossover current in the backward (forward) direction reaches the local minimum (maximum) around $t=0.6$, because the V-I curves are highly nonreciprocal there. Second, an intriguing feature can be found in the low-current regime for $t=0.6$. As shown in Fig.~\ref{fig12}, the deviation from the scaling law $V\propto I^3_s$ is evident at relatively low currents. In particular, we see the large deviation for low currents in the backward direction. We note again that the helical SC states show a crossover near the reduced temperature $t=0.6$. The fate of the nonreciprocal features around the crossover line at low temperatures near $T=0$ is also an interesting issue but beyond the scope of this paper.

\subsection{Finite Size Effect\label{VIII D}}

Let us discuss the finite size effect on the V-I characteristics. Following the spirit of the HN theory, the effect is invoked by setting the upper limit of critical separation $r_c$ in Eq.~(\ref{xc}) to be the width of the sample, leading to the restriction of the critical scale factor $l_c$. In the original HN theory, only the information of the zero current states at the zero field BKT transition temperature is considered, and the finite size effect results in the appearance of infinitesimally linear resistivity at extremely low currents. Here, we extend the theory in the same way as we did for the nonlinear V-I characteristics: Considering the quantities as functions of the current $I_s$ obtained from the mean-field theory. For example, the current dependence of the superfluid stiffness is taken into account. Based on our approach, the unexpected electric current dependence of the resistivity due to the finite size effect is shown below.

\begin{figure}[b]

\captionsetup{font=normal,justification=raggedright,singlelinecheck=false}

\begin{minipage}{0.49\hsize}
\subfloat[\label{fig13a}]{\includegraphics[width=4.2cm]{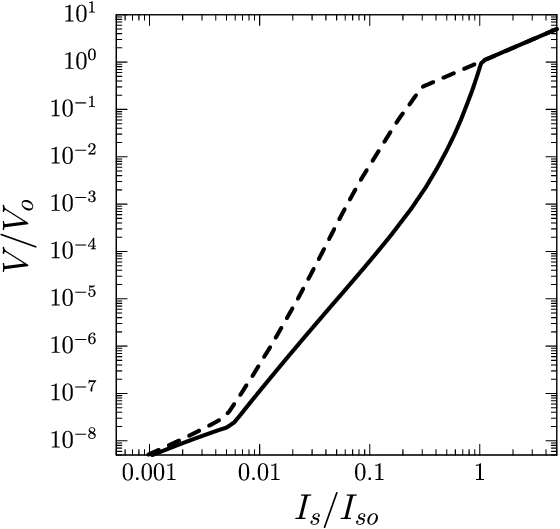}}
\end{minipage}
\begin{minipage}{0.49\hsize}
\subfloat[\label{fig13b}]{\includegraphics[width=4.2cm]{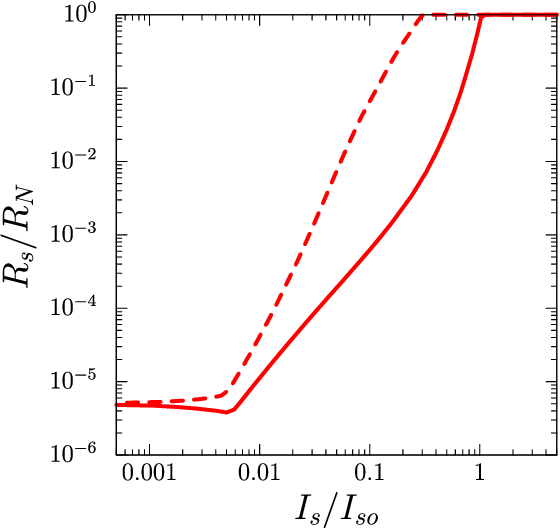}}
\end{minipage}

\captionsetup{font=normal,justification=raggedright,singlelinecheck=false}
\caption{(a) V-I characteristics and (b) the resistance curves at the BKT transition temperature $t=0.6$ in the magnetic field when the finite size effect is considered. The solid and dashed lines show the curves for the forward and backward currents, respectively.}

\label{fig13}
\end{figure}

For demonstration, we adopt $\ln \, (L/\xi_c) = 6$ in the numerical calculation, where $L$ is the system size and $\xi_c$ is the coherence length of the zero current state at the zero field BKT transition temperature. Note that a typical value of $\ln\, (L/\xi_c) \approx 9$ has long been used in the analysis of the experimental results of the BKT transition in indium and indium oxide superconducting films \cite{Fiory}. The results are shown in Fig.~\ref{fig13}, where the parameters are the same as Fig.~\ref{fig11b}. As shown in Fig.~\ref{fig13}, although the voltage increases with the current in both directions, the resistance in the forward direction decreases with the current in the extremely low current regime where the finite size effect dominates the resistivity. As a consequence, the V-I characteristics show a non-monotonic behavior. In our calculations, the non-monotonic behavior is observed only at finite magnetic fields and is enhanced around the crossover of helical SC states. To understand this behavior, we analyzed the numerical data and found that in the finite-size effect regime, $\exp\, (-2l_c\sqrt{|C_0|})$ in Eq.~(\ref{yl1}) is the dominant contribution in Eq.~(\ref{Rsf}), and the critical scale factor $l_c$ is approximately constant. Therefore, the behavior of the bare effective superfluid stiffness $J_B$ is reflected through the $C_0$ term. As shown in Fig.~\ref{fig6c}, the bare effective superfluid stiffness $J_B$ can increase with increasing the forward current due to the difference between $q_0$ and $q_J$ [see Sec.~\ref{IV C} for the definitions]. In this way, the decrease of resistivity in the regime dominated by finite size effects comes from the separation between $q_0$ and $q_J$.

The negative (positive) differential resistance in the forward (backward) direction can be regarded as an analog of the magnetochiral anisotropy~\cite{Rikken} which is defined by the correction to the linear resistivity by the current. In this view, the finite size effect on the BKT transition causes a giant magnetochiral anisotropy. The ratio of the resistivity reaches $R_s^+/R_s^- \approx 2$ at a current $I_s/I_{s0} \approx 5\times10^{-3}$.

\section{Summary and Discussions\label{IX}}

In summary, we have studied the BKT phase diagram and the nonlinear V-I characteristics of finite-momentum helical states in the purely 2D Rashba superconductors coupled with in-plane Zeeman fields. The theory of V-I characteristics was extended such that the momentum dependence of the SC states is considered. Three issues have been addressed. The first issue is to clarify the action for analyzing the BKT transition. We obtain at the first stage the effective action of vortex fluctuation including the static Gaussian amplitude fluctuation, which is a kind of the 2D continuum XY action. In this effective action, the matrix elements of the superfluid stiffness are equivalent to the second-order partial derivatives of the mean-field free energy density as a function of the Cooper pair momentum. The equivalence should be satisfied as asserted in Ref.~\cite{Liang}, and our result shows a crucial role of the coupling terms between the amplitude and phase modes in the action of helical superconductors.

The second issue is the phase diagrams of the model. We mainly investigated two subjects: the deviation of the mean-field critical magnetic field from the BKT transition one, $\Delta h$, and the separation between $q_0$ and $q_J$, the Cooper pair momenta which maximize the absolute value of condensation energy and the bare effective superfluid stiffness, respectively. The condensation energy and bare effective superfluid stiffness take a role in determining the BKT transition line. At the zero-field BKT transition temperature $T_{\rm c}$, we find that $\Delta h$ can be $30\%$ of the 2D Pauli limit field if the vortex core energy is considered, and half of that if the vortex core energy is neglected. We also find that $\Delta h$ and the difference between $q_0$ and $q_J$ exhibit non-monotonic behaviors, resulting from the crossover of helical SC states where $q_0$ drastically increases with the magnetic field accompanied by the appearance of metastable states.

The last issue is to elucidate the nonlinear resistivity of purely 2D Rashba superconductors. The HN theory was extended by including the dependence of superfluid stiffness and supercurrent on the Cooper pair momentum. These quantities are calculated from the mean-field theory of the helical superconductor. The extended HN theory gives the following results. In the vicinity of the BKT transition at zero magnetic field, our approach based on the picture of vortex-antivortex pair dissociation can generate the V-I characteristics which are consistent with some recent experimental works \cite{Venditti, Lin, Saito, Hua,  Weitzel} better than the original HN theory. This finding suggests that the deviation from the HN scaling law may be explained by our theory if the heating effect~\cite{Weitzel} is negligible. At finite magnetic fields, we found that the V-I characteristics show a nonreciprocity, in which voltages in the forward and backward directions of currents become different. We also found that the nonreciprocity in the V-I characteristics is strongly enhanced around the crossover of helical SC states. The nonreciprocal V-I characteristics are an analog of the SD effect in purely 2D superconductors, where the critical current can not be defined. Instead of the nonreciprocal critical current, a highly nonlinear and nonreciprocal resistivity appears. 

Along the BKT transition line, the crossover currents, at which the BKT critical state crossovers to the normal state, show non-monotonic behavior as a function of the magnetic field and temperature. The crossover current is an analog of the critical current in the mean-field superconductors and nonreciprocal, in particular, around the crossover line of helical SC states. In this way, we can observe an analog of the SD effect in purely 2D superconductors.

Finally, our study of the finite size effect has revealed that the nonlinear resistivity can decrease with increasing current in a finite magnetic field. This behavior can be regarded as a giant magnetochiral anisotropy, reflecting the strong nonlinearity of the BKT critical states. The microscopic origin is attributed to the difference between $q_0$ and $q_J$, the two characteristic Cooper pair momenta that maximize the condensation energy and the superfluid weight.

Some issues remain at present. Let us take them up for three topics. First, as emphasized in the main text, the expression for calculating the nonlinear resistivity is formulated for a dirty superconductor. Thus, when it is used to calculate the nonlinear resistivity in a clean system, the resistivity may be overestimated. Therefore, it is desirable to estimate the nonlinear resistivity microscopically for various strengths of disorders. Since the nonlinear resistivity in the BKT critical region should reflect the results of the mean-field theory, we may see nonreciprocity similar to the SD effect of mean-field bulk superconductors. It has been shown that the SD effect in the clean limit of the isotropic system does not change the sign near the transition temperature as increasing the magnetic field~\cite{Ilic}~\footnote{A sizable SD effect near the transition temperature arises from the anisotropy in the electronic band dispersion, and the sign change occurs even in the clean limit~\cite{Daido1,Daido2}.}, but it can change the sign in the presence of disorders. From this result, it is expected that the nonlinear resistivity shows opposite nonreciprocity between the low and high magnetic field regions when we consider a model with disorders.

Second, the phase diagram deep below the BKT transition temperature has not been resolved in the helical superconductors. In the mean-field phase diagram of the attractive Rashba-Hubbard model, the crossover of helical SC states changes to a first-order transition at low temperatures, and the Cooper pair momentum discontinuously changes there~\cite{Daido1}. The fate of the first-order transition in the quasi-long-range ordered state of purely 2D superconductors remains elusive. A natural question is whether the first-order transition that separates the mean-field phase diagram into the low- and high-magnetic field phases still exists in purely 2D superconductors.

Third, the effect of high-order terms of the Kosterlitz renormalization group equations on the properties of purely 2D superconductors is not clarified in this paper. A recent work~\cite{Maccari} has studied the renormalization flow of the XY model by using the functional renormalization group method, by which higher-order terms can be taken into account. This method may solve the second issue mentioned in the previous paragraph. Therefore, it is interesting to combine the method with our approach to study the nonlinear diode effect.

In addition to the present classical scheme on calculating the free-vortex generating rate, there is a quantum path integral formulation including quantum dissipation effect \cite{Iengo}. In the latter, the correction term to the nonlinear resistivity at temperatures below the zero field BKT transition temperature $T_{\rm c}$ has been derived, and the voltage $V$ and current $I$ are related by $V=c_1I^{a} + c_2 I^{2+a}$, where $c_1,c_2$ are coefficients and $a\rightarrow 3$ for $T\rightarrow T_{\rm c}$. Since the centrosymmetric model was studied in Ref.~\cite{Iengo}, the $I^{1+a}$ term vanishes in the above expression. It would be interesting to apply the formulation in Ref.~\cite{Iengo} to noncentrosymmetric superconductors. At present it is unclear whether the same results as ours will be obtained at the lowest order.

To study purely 2D superconductors, calculating the superfluid stiffness is inevitable as we have done in this paper. The formulation of the superfluid stiffness is also important for 3D superconductors. According to Ref.~\cite{Fisher}, as the current increases, the Meissner phase of bulk superconductors can be broken by the generation of isolated vortex loops. To calculate the free energy of vortex loops, the information on the superfluid stiffness is needed~\cite{Langer}. Thus, when the critical current is dominated by the generation of vortex loops, the superfluid stiffness plays an essential role. In noncentrosymmetric superconductors, the SD effect is expected to arise from this mechanism of the critical current. The nonreciprocity of the vortex loop generation is an interesting topic for future studies.

The superconducting properties under the supercurrent, such as the SD effect and the current-induced optical responses, are topics of recent interest. We hope that the formulation for current-carrying superconductors studied in this paper will be useful in a wide range of topics.

\section*{Acknowledgements}

We acknowledge useful discussions with A.~Daido, T.~Kitamura, and R.~Ikeda.
This work was supported by JSPS KAKENHI Grant Numbers JP21K18145, JP22H01181, JP22H04933, JP23K17353, JP24K21530, JP24H00007.

\appendix*

\def\thesection{\Alph{section}}

\section{Effective action\label{AppendixA}}

\subsection{Regularization of the integration associated with the phase fluctuations\label{AppendixA1}}

In Eq.~(\ref{eq29}), the Fourier components of the spin wave part of the phase fluctuation $\theta_s(x)$, namely $\theta_{qs}$, should be bounded in some ranges, namely $\theta_{qs}\in[-\Pi,\Pi]$, where $\Pi$ is a positive real number. This is because $\theta_s(x)$ is limited on the interval $[-\pi,\pi]$ with the introduction of the Hubbard-Stratonovich transformation in Eq.~(\ref{eq3}). To evaluate the integration, we introduce a regularizing function $f_R$ by which the integration bound can be extended to infinity while principally keeping the result of the integration. Here we chose $f_R(q) = \frac{1}{\pi \Pi_R} \exp\lr{-\theta_{qs}\theta_{-qs}/\Pi_R}$, where $\Pi_R$ is a positive real number and does not need to be equal to $\Pi$. We will see soon that the effective action for transverse gauge fields does not depend on the choice of $\Pi_R$. Inserting the regularizing function into Eq.~(\ref{eq29}), we obtain the partition function for the mode with momentum $q$, $Z(q)$, as follows:
\[   Z(q) = \int \frac{d\theta_{qs}d\theta_{-qs}}{\pi\Pi_R}\, e^{-S_q(A) - \frac{\theta_{qs}\theta_{-qs}}{\Pi_R}},    \tag{A1} \label{a1}  \]
\[S_q(A) =\tilde{J}_{\mu\nu}(iq_{\mu}\theta_{qs} + \hat{A}_{q\mu})(-iq_{\nu}\theta_{-qs} + \hat{A}_{-q\nu}),      \tag{A2}   \label{a2}  \] 
where $\tilde{J}_{\mu\nu}$ is the $\mu\nu$ component of the total superfluid stiffness, $\hat{A}_{q\mu}$ is a Fourier component of $\hat{A}_{\mu}(x) = \partial_{\mu}\theta_{v}(x) + 2e A_{\mu}(x)$, and $\theta_{v}$ denotes the vortex part of the phase fluctuation. 

\subsection{Derivation of effective action for transverse gauge fluctuations\label{AppendixA2}}

In the following, the aim is to obtain the Gaussian effective action of the transverse gauge fields. To do this, we have to integrate out $\theta_s$ and the longitudinal gauge field defined below from the action $S_q(A)$. We first perform the former or Eq.~(\ref{a1}). There are two ways to tackle it. One is the method introduced at the beginning of Sec.~\ref{IV A}, in which the gauge transformation is performed to include all values of $\theta_{qs}$ in $A_{q\mu}$, called \textit{procedure I}. Following this procedure, we have
\[ \int \! D(A_q,A_{-q}) \, Z(q) = \int \! D(A_q,A_{-q}) \, e^{-\tilde{J}_{\mu\nu}\hat{A}_{q\mu}\hat{A}_{q\nu} },   \tag{A3}  \label{a3}    \]
where $D(A_q,A_{-q})$ is a Fourier component of the measure $D[A]$ in Eq.~(\ref{eq29}). It is obvious from the above equation that the dependence on $\Pi_R$ has already vanished in this procedure. To proceed further, let us consider $A^L_{q\mu}=(\hat{q}_{\nu}A_{q\nu})\,\hat{q}_{\mu}$ with $\hat{q}_{\mu}=q_{\mu}/\sqrt{q^2}$ and $q^2=q_{\mu}q_{\mu}$ and $A^T_{q\mu}=A_{q\mu}-A^L_{q\mu}$ for the longitudinal and transverse parts of $A_{q\mu}$, respectively. To diminish the ambiguity of the $q=0$ mode, we here do not classify it as a longitudinal field. Performing the integration with respect to the longitudinal components in Eq.~(\ref{a3}), we obtain the following expressions:
 \begin{align*} 
 \int & D(A_q,A_{-q})  \, Z(q) \\
 &\quad = \int D(A^T_q,A^T_{-q}) \, e^{-K_{\mu\nu}  \hat{A}^T_{q\mu} \hat{A}^T_{-q\nu} - \ln \Gamma_q},     \tag{A4}  \label{a4}         
\end{align*}
\[     K_{\mu\nu} = \tilde{J}_{\mu\nu}   - \frac{ q_{\alpha}  \tilde{J}_{\mu\alpha} q_{\beta} \tilde{J}_{\beta\nu}     }{ q_{\alpha}q_{\beta}    \tilde{J}_{\alpha\beta}     },      \tag{A5} \label{a5}            \]
\[      \Gamma_q = \hat{q}_{\mu}\tilde{J}_{\mu\nu}\hat{q}_{\nu},    \tag{A6} \label{a6}  \]
where $\hat{A}^L_{q\mu}=(\hat{q}_{\nu}\hat{A}_{q\nu})\,\hat{q}_{\mu}$, $\hat{A}^T_{q\mu}=\hat{A}_{q\mu} - \hat{A}^L_{q\mu}$, and $K_{\mu\nu}$ is the gauge-invariant superfluid stiffness or the EM response function. The Jacobian accompanied by the change in the variables depends only on $q$ and is absorbed in the integral measure $D(A^T_q,A^T_{-q})$. We would like to remind that the contribution from the uniform and static mode with $\bm{q}=\omega_n=0$ has been removed due to the convention made above. Here, $\Gamma_q$ denotes the dispersion relation of the collective mode associated with the longitudinal gauge fields, which depends on the gauge fixing scheme. We will discuss this later. 

When the momentum dependence of $\tilde{J}_{\mu\nu}$ is taken into account, we can rewrite $\tilde{J}_{\mu\nu}$ as $\tilde{J}_{\mu\nu}(q)$. We can show that this replacement results in the expressions in Eqs.~(\ref{a5}) and (\ref{a6}) with $\tilde{J}_{\mu\nu}$ replaced by $\tilde{J}_{\mu\nu}(q)$.

The above procedure of integrating out the longitudinal gauge field can be regarded as a gauge fixing. With the choice of the longitudinal part used above, the \textit{Lorenz gauge fixing} is indeed implied. Strictly speaking, to fix the gauge condition on the path integral, one may use the Faddeev-Popov method. By using this method, the integration of the longitudinal gauge field can be well defined even in a normal state, in which the integration formally diverges. In the SC phase, as shown above, the integration of the longitudinal gauge field converges, and the Faddeev-Popov method seems to be unnecessary. The Faddeev-Popov method on the path integral as Eq.~(\ref{a3}) can be found in Ref.~\cite{Harada}. Following this literature, we obtain the same results as Eqs.~(\ref{a4}), (\ref{a5}), and (\ref{a6}).

The other way to deal with Eq.~(\ref{a1}), called \textit{procedure II}, is to integrate $\theta_{qs}$ and $\theta_{-qs}$ while keeping $\hat{A}_{q\mu}$ and $\hat{A}_{-q\nu}$ constant. Following this procedure, we obtain $Z(q) = \exp\lr{-S_q(A)}$ with
\begin{align*}     
 S_q(A)  &=   \tilde{J}_{\mu\nu} \hat{A}_{q\mu} \hat{A}_{-q\nu}  - \frac{  (\hat{q}_{\mu} \tilde{J}_{\mu\nu} \hat{A}_{q\nu} ) ( \hat{q}_{\beta}\tilde{J}_{\alpha\beta}\hat{A}_{-q\alpha} ) }{\Pi_R + (\Gamma_q \, q^2)^{-1} } \\
&\quad +  \ln\lr{ 1 + \Pi_R\Gamma_q \, q^2   }. \tag{A7} \label{a7}  
\end{align*}
Thus, in contrast to the previous method, a massive collective mode with the mass proportional to the cutoff $\Pi^{-1}_R$ appears after integrating the phase fluctuation. If $\Pi^{-1}_R$ is taken to infinity at this stage, the massive mode becomes the Goldstone mode, and the controversy as mentioned in Sec.~\ref{IV A} arises. However, the problem can be solved by performing the integration of the above equation with respect to the longitudinal gauge fields while keeping $\Pi_R$ finite. With some elementary algebras, we can obtain Eq.~(\ref{a4}) again after integrating out the longitudinal gauge fields. In this way, the two procedures I and II mentioned above coincide with each other when the longitudinal gauge fields are integrated from the action. 

Let us turn back to the discussion of the collective mode associated with Eq.~(\ref{a6}). If the dependence of superfluid stiffness $J_{\mu\nu}$ and $\tilde{J}_{\mu\nu}$ on $q_{\mu}$ can be neglected, this mode is massless and has to be interpreted as the Nambu-Goldstone mode \cite{Nambu}. Due to the breaking of parity and time-reversal symmetries, the spectrum of this mode is directional dependent. To make this mode massive, we have to invoke the Hamiltonian for the Coulomb interaction $V(\bm{q})$ in the microscopic model Eq.~(\ref{eq2}). Assuming $V(\bm{q}) = \alpha_{\mathrm{c}}/\bm{q}^2$, we can show that the equation 
\[\lim{\bm{q}}{0}K_{00}(\omega,\bm{q})V(\bm{q}) =1, \tag{A8} \label{a8} \] 
determines the mode's energy in the limit $\bm{q}\rightarrow 0$, where we have already performed the analytic continuation $i\omega_n\rightarrow \omega$ from a bosonic Matsubara frequency $\omega_n$ to a real frequency $\omega$. In the conventional theories, in which only the $s$-wave attractive interaction and the Coulomb interaction are included and renormalization due to the amplitude fluctuation in Eq.~(\ref{eq26}) is neglected, the above equation shows the spectrum of isotropic plasmon mode. In our model, however, the dependence of this mode on various parameters becomes nontrivial and needs to be further investigated.

Let us end the section by discussing the results when the \textit{Coulomb gauge fixing} is imposed. In this gauge, the longitudinal gauge field is restricted only in the spatial coordinates as $A^L_{i} = (\hat{q}_{j}A_{qj})\hat{q}_i$, where $\hat{q}_i$ is a spatial component of $\hat{q}_{\mu}$. Then, the electric field part of the pure EM action $S_{\mathrm{E}}[A]$, which is part of $S_{\mathrm{a}}[A]$, should be carefully treated because this part is not invariant under the transformation $A_{i}(x) \rightarrow A_{i}(x) + \partial_i \chi(x)$, where $\chi$ is a smooth function. This part of the action has the form 
\[        S_{\mathrm{E}}[A] = \frac{1}{2} \alpha^2_{\rm c} \int d^3x\, (\partial_i A_0 - \partial_0 A_i)^2.       \tag{A9} \label{a9} \]
We first look at the EM responses. For simplicity, we only work on procedure I. After integrating the longitudinal part in this gauge, the EM response function in the Coulomb gauge, $K^{\rm c}_{ij}$, has the form
\[   K^{\mathrm{c}}_{ij} = \tilde{J}^{\rm c}_{ij}   - \frac{  \hat{q}_l \tilde{J}^{\rm c}_{il} \hat{q}_m \tilde{J}^{\rm c}_{mj} }{\hat{q}_l\tilde{J}^{\rm c}_{lm}\hat{q}_m },       \tag{A10} \label{a10}            \]
where $\tilde{J}^{\rm c}_{ij}$ is defined as follows:
\begin{align*}
\tilde{J}^{\rm c}_{ij} &= \tilde{J}_{ij} + \alpha^2_{\rm c} \, \omega^2_n \delta_{ij}, \tag{A11a} \label{a11a} \\
\tilde{J}^{\rm c}_{0i} &= \tilde{J}^{\rm c}_{i0} =\tilde{J}_{0i} + \alpha^2_{\rm c} \,\omega_n q_i , \tag{A11b} \label{a11b} \\
\tilde{J}^{\rm c}_{00} &= \tilde{J}_{00} + \alpha^2_{\rm c} \, \bm{q}^2. \tag{A11c} \label{a11c}
\end{align*}
The dispersion relation of the longitudinal gauge field now takes the form $\Gamma^{\rm c}_q=\hat{q}_i \tilde{J}^{\rm c}_{ij} \hat{q}_j $. Performing the analytic continuation $i\omega_n\rightarrow\omega$, this equation is reduced to the same form as Eq.~(\ref{a8}). Therefore, the plasmon mode is naturally implied in the Coulomb gauge without adding the Coulomb Hamiltonian.

 \def\thesection{\Alph{section}}

\section{Vortex core radius dependence of the factor $\exp(-2l_c\sqrt{|C_0|})$ at low temperatures\label{AppendixB}}

At sufficiently low temperatures, $K_B=J_B/T$ grows large such that $-\sqrt{|C_0|}\simeq X_0 = 1-\frac{\pi}{2}K_B$ [See the definitions of $J_B$ and $C_0$ below Eqs.~(\ref{Svo}) and (\ref{BKTinr}), respectively]. From Eq.~(\ref{xc}) we obtain the expression 
\[\exp\, (l_c) = \eta J(l_c)/(\xi j_{s2}),    \tag{B1} \label{b1} \] 
where $\eta=1$ $(\eta=1/0.74)$ when the vortex core radius is chosen to be the BCS (GL) coherence length. Since $C_0$ depends only on $K_B$ at sufficiently low temperatures as explained above,  we find $X(l)\simeq -\sqrt{|C_0|}$ from Eq.~\eqref{XX1}, meaning that the dependence of $X$ on the scale factor $l$ can be neglected. Taking these into account, we notice that the r.h.s of Eq.~(\ref{b1}) is just a constant. Thus, we can conclude that the critical scale factor $l_c$ increases with $\eta$ at sufficiently low temperatures. Therefore, for a given supercurrent and magnetic field, at sufficiently low temperatures, the factor $\exp(-2l_c\sqrt{|C_0|})$ is smaller when the GL coherence length is assumed to be the vortex core radius than when the the BCS coherence length is assumed. The reduction of this factor when we adopt the GL coherence length as a vortex core radius is essential for the results in Sec.~\ref{VIII B}.


\begin{thebibliography}{99}


\bibitem{Daido1} A. Daido, Y. Ikeda, and Y. Yanase, Intrinsic Superconducting Diode Effect, Phys. Rev. Lett. 128, 037001 (2022).

\bibitem{Yuan} N. F. Q. Yuan and L. Fu, Supercurrent diode effect and finite-momentum superconductors, Proc. Nati. Acad. Sci. USA 119, e2119548119 (2022)

\bibitem{He-Nagaosa}
J. J. He, Y. Tanaka, and N. Nagaosa, A phenomenological theory of superconductor diodes,
New J. Phys. 24, 053014 (2022).

\bibitem{Daido2}  A. Daido and Y. Yanase, Superconducting diode effect and nonreciprocal transition lines, Phys. Rev. B 106, 205206 (2022).

\bibitem{Ilic} S. Ili\'{c} and F. S. Bergeret, Theory of the Supercurrent Diode Effect in Rashba Superconductors with Arbitrary Disorder, Phys. Rev. Lett. 128, 177001 (2022).

\bibitem{Ando} F. Ando, Y. Miyasaki, T. Li, J. Ishizuka, T. Arakawa, Y. Shiota, T. Moriyama, Y. Yanase, and T. Ono, Observation of superconducting diode effect, Nature 584, 373 (2020).

\bibitem{Bauriedl} L. Bauriedl, C. B\"{a}uml, L. Fuchs, C. Baumgartner, N. Paulik, J. M. Bauer, KQ. Lin, J. M. Lupton, T. Taniguchi, K. Watanabe \textit{et al.}, Supercurrent diode effect and magnetochiral anisotropy in few-layer NbSe2, Nat Commun 13, 4266 (2022).


\bibitem{Narita} H. Narita, J. Ishizuka, D. Kan, Y. Shimakawa, Y. Yanase, and T. Ono, Magnetization Control of Zero-Field Intrinsic Superconducting Diode Effect, Adv. Mater. 2023, 2304083 (2023).

\bibitem{Nagaosa-Yanase}
Naoto Nagaosa and Youichi Yanase, 
Nonreciprocal Transport and Optical Phenomena in Quantum Materials, 
Annual Review of Condensed Matter Physics, 15, 63 (2024). 


\bibitem{Aoyama} K. Aoyama, Stripe order and diode effect in two-dimensional Rashba superconductors, Phys. Rev. B 109, 024516 (2024).

\bibitem{Nunchot} N. Nunchot and Y. Yanase, Chiral superconducting diode effect by Dzyaloshinsky-Moriya interaction, Phys. Rev. B 109, 054508 (2024).

\bibitem{Kosterlitz1} J. M. Kosterlitz and D. J. Thouless, Ordering, metastability and phase transitions in two-dimensional systems, J. Phys. C: Solid State Phys. 6, 1181 (1973).
\bibitem{Kosterlitz2} J. M. Kosterlitz, The critical properties of the two-dimensional xy model, J. Phys. C: Solid State Phys. 7, 1046 (1974).

\bibitem{Nelson} D. R. Nelson and J. M. Kosterlitz, Universal Jump in the Superfluid Density of Two-Dimensional Superfluids, Phys. Rev. Lett. 39, 1201 (1977).

\bibitem{Pearl} J. Pearl, Current distribution in superconducting films carrying quantized fluxoids, Appl. Phys. Lett. 5, 65-66 (1964).

\bibitem{Beasley} M. R. Beasley, J. E. Mooji, and T. P. Orlando, Possibility of Vortex-Antivortex Pair Dissociation in Two-Dimensional Superconductors, Phys. Rev. Lett. 42, 1165 (1979).

\bibitem{Gong} M. Gong, G. Chen, S. Jia, and C. Zhang, Searching for Majorana Fermions in 2D Spin-Orbit Coupled Fermi Superfluids at Finite Temperature, Phys. Rev. Lett. 109, 105302 (2012).

\bibitem{Yin} S. Yin, J.-P. Martikainen, and P. T\"{o}rm\"{a}, Fulde-Ferrell states and Berezinskii-Kosterlitz-Thouless phase transition in two-dimensional imbalanced Fermi gases, Phys. Rev. B 89, 014507 (2014).

\bibitem{Xu2} Y. Xu and C. Zhang, Berezinskii-Kosterlitz-Thouless Phase Transition in 2D Spin-Orbit-Coupled
 Fulde-Ferrell Superfluids, Phys. Rev. Lett. 114, 110401 (2015).

\bibitem{Aitchison} I. J. R. Aitchison, P. Ao, D. J. Thouless, and X. -M. Zhu, Effective Lagrangians for BCS superconductors at $T=0$, Phys. Rev. B 51, 6531 (1995).

\bibitem{Botelho} S. S. Botelho and C. A. R. S\'{a} de Melo, Vortex-Antivortex Lattice in Ultracold Fermionic Gases, Phys. Rev. Lett. 96, 040404 (2006).

\bibitem{Diener} R. B. Diener, R. Sensarma, and M. Randeria, Quantum fluctuations in the superfluid state of the BEC-BCS crossover, Phys. Rev. A 77, 023626 (2008).


\bibitem{Salasnich} L. Salasnich and F. Toigo, Zero-point energy of ultracold atoms, Phys. Rep. 640, 1 (2016).

\bibitem{He} L. He, H. L\"{u}, G. Cao, H. Hu, and X. J. Liu, Quantum fluctuations in the BCS-BEC crossover of two-dimensional Fermi gases, Phys. Rev. A 92, 023620 (2015).

\bibitem{Bighin} G. Bighin and L. Salasnich, Finite-temperature quantum fluctuations in two-dimensional Fermi superfluids, Phys. Rev. B 93, 014519 (2016).


\bibitem{Anderson} P. W. Anderson, Coherent Excited States in the Theory of Superconductivity: Invariance and the Meissner Effect, Phys. Rev. 110, 827 (1958); P. W. Anderson, Plasmons, Gauge Invariance, and Mass, Phys. Rev. 130, 439 (1963).


\bibitem{Ambegaokar} V. Ambegaokar, B. I. Halperin, D. R. Nelson, and E. D. Siggia, Dynamics of superfluid films, Phys. Rev. B, 21, 1806 (1980).

\bibitem{Halperin} B. I. Halperin and D. R. Nelson, Resistive transition in superconducting films, J. Low Temp. Phys. 36, 599-616 (1979).

\bibitem{Hebard} A. F. Hebard and A. T. Fiory, Critical-Exponent Measurements of a Two-Dimensional Superconductor, Phys. Rev. Lett. 50, 1603 (1983).

\bibitem{Epstein} A. M. Kadin, K. Epstein, and A. M. Goldman, Renormalization and the Kosterlitz-Thouless transition in a two-dimensional superconductor, Phys. Rev. B 27, 6691 (1983).

\bibitem{Fiory} A. T. Fiory, A. F. Hebard, and W.I. Glaberson, Superconducting phase transitions in indium/indium-oxide thin-film composites, Phys. Rev. B 28, 5075 (1983).

\bibitem{Garland} J. C. Garland and H. J. Lee, Influence of a magnetic field on the two-dimensional phase transition in thin-film superconductors, Phys. Rev. B 36, 3638 (1987).

\bibitem{Lin} Z. Lin, C. Mei, L. Wei \textit{et al.}, Quasi-two-dimensional superconductivity in FeSe$_{0.3}$Te$_{0.7}$ thin films and electric-field modulation of superconducting transition, Sci Rep 5, 14133 (2015).

\bibitem{Venditti} G. Venditti, J. Biscaras, S. Hurand, N. Bergeal, J. Lesueur, A. Dogra, R. C. Budhani, M. Mondal, J. Jesudasan, P. Raychaudhuri \textit{et al.}, Nonlinear I-V characteristics of two-dimensional superconductors: Berezinskii-Kosterlitz-Thouless physics versus inhomogeneity, Phys. Rev. B 100, 064506 (2019).

\bibitem{Saito} Y. Saito, Y. M. Itahashi, T. Nojima, and Y. Iwasa, Dynamical vortex phase diagram of two-dimensional superconductivity in gated MoS$_2$, Phys. Rev. Materials 4, 074003 (2020).

\bibitem{Hua}  X. Hua, F. Meng, Z. Huang \textit{et al.}, Tunable two-dimensional superconductivity and spin-orbit coupling at the EuO/KTaO3(110) interface, npj Quantum Mater. 7, 97 (2022). 

\bibitem{Weitzel} A. Weitzel, L. Pfaffinger, I. Maccari, K. Kronfeldner, T. Huber, L. Fuchs, J. Mallord, S. Linzen, E. Il'ichev, N. Paradiso, and C. Strunk, Sharpness of the Berezinskii-Kosterlitz-Thouless Transition in Disordered NbN Films, Phys. Rev. Lett. 131, 186002 (2023).

\bibitem{Liu}  C. Liu, X. Zhou, D. Hong \textit{et al.}, Tunable superconductivity and its origin at KTaO$_3$ interfaces, Nat Commun 14, 951 (2023).




\bibitem{Hoshino} S. Hoshino, R. Wakatsuki, K. Hamamoto, and N. Nagaosa, Nonreciprocal charge transport in two-dimensional noncentrosymmetric superconductors, Phys. Rev. B. 98, 054510 (2018).


\bibitem{Qiu} D. Qiu, C. Gong, S. Wang, M. Zhang, C. Yang, X. Wang, and J. Xiong, Recent advances in 2D superconductors, Adv. Mater. 33, 2006124 (2021).


\bibitem{DFisher} D. S. Fisher, Flux-lattice melting in thin-film superconductors, Phys. Rev. B 22, 1190 (1980).

\bibitem{Liang} L. Liang, T. I. Vanhala, S. Peotta, T. Siro, A. Harju, and P. T\"{o}rm\"{a}, Band geometry, Berry curvature, and superfluid weight, Phys. Rev. B 95, 024515 (2017).

\bibitem{Samokhin} K. V. Samokhin and B. P. Truong, Current-carrying states in Fulde-Ferrell-Larkin-Ovchinnikov superconductors, Phys. Rev. B 96, 214501 (2017).

\bibitem{Boyack} R. Boyack, B. M. Anderson, C.-T. Wu, and K. Levin, Gauge-invariant theories of linear response for strongly correlated superconductors, Phys. Rev. B 94, 094508 (2016).


\bibitem{Coleman} P. Coleman, \textit{Introduction to Many-Body Physics}  (Cambridge University Press, Cambridge, 2015).


\bibitem{Altland} A. Altland and B. Simons, \textit{Condensed Matter Field Theory}, 2nd ed. (Cambridge University Press, New York, 2010).

\bibitem{Nambu} Y. Nambu, Quasi-Particles and Gauge Invariance in the Theory of Superconductivity, Phys. Rev. 117, 648 (1960).


\bibitem{Mondal} M. Mondal, S. Kumar, M. Chand, A. Kamlapure, G. Saraswat, G. Seibold, L. Benfatto, and P. Raychaudhuri, Role of the Vortex-Core Energy on the Berezinskii-Kosterlitz-Thouless Transition in Thin Films of NbN, Phys. Rev. Lett. 107, 217003 (2011).

\bibitem{Kramer} L. Kramer and W. Pesch, Core Structure and Low-Energy Spectrum of Isolated Vortex Lines in Clean Superconductors at $T\ll T_c$, Z. Physik 269, 59-64 (1974).


\bibitem{Larkvv} A. Larkin and A. A. Varlamov, \textit{Theory of fluctuations in superconductors} (Clarendon Press, Oxford, 2005).


\bibitem{Agterberg} D. F. Agterberg and R. P. Kaur, Magnetic-field-induced helical and stripe phases in Rashba superconductors, Phys. Rev. B 75, 064511 (2007).

\bibitem{Berg} E. Berg, E. Fradkin, and S. A. Kivelson, Charge-4e superconductivity from pair-density-wave order in certain high-temperature superconductors, Nature Physics 5, 830-833 (2009).

\bibitem{Clogston} A. M. Clogston, Upper Limit for the Critical Field in Hard Superconductors, Phys. Rev. Lett. 9, 266 (1962).


\bibitem{Daido_para} Akito Daido, Youichi Yanase, Rectification and Nonlinear Hall effect by Fluctuating Finite-momentum Cooper Pairs, Phys. Rev. Research 6, L022009 (2024).

\bibitem{Clem} J. R. Clem and K. K. Berggren, Geometry-dependent critical currents in superconducting nanocircuits, Phys. Rev. B 84, 174510 (2011).

\bibitem{Sonin} E. B. Sonin, Magnus force in superfluids and superconductors, Phys. Rev. B 55, 485 (1997).

\bibitem{Chandrasekhar} S. Chandrasekhar, Stochastic Problems in Physics and Astronomy, Rev. Mod. Phys. 15, 58 (1943).

\bibitem{Kim} Y. B. Kim and M. J. Stephen, Superconductivity In Two Parts: Volume 2, R. D. Park, ed. (Marcel Dekker, New York, 1969), Chapter 19.

\bibitem{Bardeen} J. Bardeen and M. J. Stephen, Theory of the Motion of Vortices in Superconductors, Phys. Rev. 140, A1197 (1965).

\bibitem{Kopnin} N. B. Kopnin, \textit{Theory of Nonequilibrium Superconductivity} (Oxford University Press, 2001).

\bibitem{Skvortsov} M. A. Skvortsov, D. A. Ivanov, and G. Blatter, Vortex viscosity in the moderately clean limit of layered superconductors, Phys. Rev. B 67, 014521 (2003).


\bibitem{Benfatto} L. Benfatto, C. Castellani, and T. Giamarchi, Broadening of the Berezinskii-Kosterlitz-Thouless superconducting transition by inhomogeneity and finite-size effects, Phys. Rev B 80, 214506 (2009).

\bibitem{Konig} E. J. K\"{o}nig, I. V. Protopopov, A. Levchenko, I. V. Gornyi, and A. D. Mirlin, Resistance of two-dimensional superconducting films, Phys. Rev. B 104, L100507 (2021).

\bibitem{Fetter} A. L Fetter and J. D. Walecka, \textit{Quantum Theory of Many-Particle Systems} (Dover, New York, 2003).

\bibitem{Rikken} G. L. J. A. Rikken, J. F\"{o}lling, and P. Wyder, Electrical Magnetochiral Anisotropy, Phys. Rev. Lett. 87, 236602 (2001).


\bibitem{Maccari} I. Maccari, N. Defenu, L. Benfatto, C. Castellani, and T. Enss, Interplay of spin waves and vortices in the two-dimensional XY model at small vortex-core energy, Phys. Rev. B 102, 104505 (2020).



\bibitem{Iengo} R. Iengo and G. Jug, Vortex pair production and decay of a two-dimensional supercurrent by a quantum-field-theory approach, Phys. Rev. B 52, 7537 (1995).



\bibitem{Fisher} D. S. Fisher, M. P. A. Fisher, and D. A. Huse, Thermal fluctuations, quenched disorder, phase transitions, and transport in type-II superconductors, Phys. Rev. B 43, 130 (1991).

\bibitem{Langer} J. S. Langer and M. E. Fisher, Intrinsic Critical Velocity of a Superfluid, Phys. Rev. Lett. 19, 560 (1967).


\bibitem{Harada} K. Harada and I. Tsutsui, Revealing the gauge freedom in the path-integral formalism, Prog. Theo. Phys. 78, 878 (1987).



\end{thebibliography}
\end{document}